\documentclass[iop]{emulateapj}
\usepackage{amsmath,amssymb,amstext}
\usepackage{graphicx}
\usepackage{url}
\usepackage[flushleft]{threeparttable}
\usepackage[breaklinks,colorlinks,citecolor=blue,linkcolor=magenta]{hyperref} 
\usepackage[all]{hypcap} 
\usepackage{booktabs}
\usepackage{natbib}

\shorttitle{DES QSO Selection}
\shortauthors{Tie et al.}

\begin{document}
\title{A Study of Quasar Selection in the 
Dark Energy Survey Supernova fields}
\author{
S. S.~Tie\altaffilmark{1},
P.~Martini\altaffilmark{2,1},
D.~Mudd\altaffilmark{1},
F.~Ostrovski\altaffilmark{3,4},
S.~L.~Reed\altaffilmark{3,4},
C.~Lidman\altaffilmark{5},
C.~Kochanek\altaffilmark{2,1},
T.~M.~Davis\altaffilmark{6},
R.~Sharp\altaffilmark{7},
S.~Uddin\altaffilmark{8},
A.~King\altaffilmark{6},
W.~Wester\altaffilmark{9},
B.E.~Tucker\altaffilmark{7},
D.~L.~Tucker\altaffilmark{9},
E.~Buckley-Geer\altaffilmark{9},
D.~Carollo\altaffilmark{10,11},
M.~Childress\altaffilmark{12,7},
K.~Glazebrook\altaffilmark{8},
S.R.~Hinton\altaffilmark{6},
G.~Lewis\altaffilmark{13},
E.~Macaulay\altaffilmark{6},
C.~R.~O'Neill\altaffilmark{6},
T. M. C.~Abbott\altaffilmark{14},
F.~B.~Abdalla\altaffilmark{15,16},
J.~Annis\altaffilmark{9},
A.~Benoit-L{\'e}vy\altaffilmark{17,15,18},
E.~Bertin\altaffilmark{17,18},
D.~Brooks\altaffilmark{15},
A. Carnero Rosell\altaffilmark{19,20},
M.~Carrasco~Kind\altaffilmark{21,22},
J.~Carretero\altaffilmark{23,24},
C.~E.~Cunha\altaffilmark{25},
L.~N.~da Costa\altaffilmark{19,20},
D.~L.~DePoy\altaffilmark{26},
S.~Desai\altaffilmark{27},
P.~Doel\altaffilmark{15},
T.~F.~Eifler\altaffilmark{28},
A.~E.~Evrard\altaffilmark{29,30},
D.~A.~Finley\altaffilmark{9},
B.~Flaugher\altaffilmark{9},
P.~Fosalba\altaffilmark{23},
J.~Frieman\altaffilmark{9,31},
J.~Garc\'ia-Bellido\altaffilmark{32},
E.~Gaztanaga\altaffilmark{23},
D.~W.~Gerdes\altaffilmark{30},
D.~A.~Goldstein\altaffilmark{33,34},
D.~Gruen\altaffilmark{25,35},
R.~A.~Gruendl\altaffilmark{21,22},
G.~Gutierrez\altaffilmark{9},
K.~Honscheid\altaffilmark{2,36},
D.~J.~James\altaffilmark{37,14},
K.~Kuehn\altaffilmark{5},
N.~Kuropatkin\altaffilmark{9},
M.~Lima\altaffilmark{38,19},
M.~A.~G.~Maia\altaffilmark{19,20},
J.~L.~Marshall\altaffilmark{26},
F.~Menanteau\altaffilmark{21,22},
C.~J.~Miller\altaffilmark{29,30},
R.~Miquel\altaffilmark{39,24},
R.~C.~Nichol\altaffilmark{40},
B.~Nord\altaffilmark{9},
R.~Ogando\altaffilmark{19,20},
A.~A.~Plazas\altaffilmark{28},
A.~K.~Romer\altaffilmark{41},
E.~Sanchez\altaffilmark{42},
B.~Santiago\altaffilmark{43,19},
V.~Scarpine\altaffilmark{9},
M.~Schubnell\altaffilmark{30},
I.~Sevilla-Noarbe\altaffilmark{42},
R.~C.~Smith\altaffilmark{14},
M.~Soares-Santos\altaffilmark{9},
F.~Sobreira\altaffilmark{19,44},
E.~Suchyta\altaffilmark{45},
M.~E.~C.~Swanson\altaffilmark{22},
G.~Tarle\altaffilmark{30},
D.~Thomas\altaffilmark{40},
A.~R.~Walker\altaffilmark{14}
\\ \vspace{0.2cm} (The DES Collaboration) \\
}

\affil{Author affiliations are listed at the end of this paper.}

\begin{abstract}
We present a study of quasar selection using the Dark Energy Survey (DES) supernova fields. We used a quasar catalog from an overlapping portion of the SDSS Stripe 82 region to quantify the completeness and efficiency of selection methods involving color, probabilistic modeling, variability, and combinations of color/probabilistic modeling with variability. In all cases, we only considered objects that appear as point sources in the DES images. We examine color selection methods based on the WISE mid-IR $W1-W2$ color, a mixture of WISE and DES colors ($g-i$ and $i-W1$) and a mixture of VHS and DES colors ($g-i$ and $i-K$). For probabilistic quasar selection, we used \texttt{XDQSOz}, an algorithm that employs an empirical multi-wavelength flux model of quasars to assign quasar probabilities. Our variability selection uses the multi-band $\chi^{2}$-probability that sources are constant in the DES Year 1 $griz$-band light curves. The completeness and efficiency are calculated relative to an underlying sample of point sources that are detected in the required selection bands and pass our data quality and photometric error cuts. We conduct our analyses at two magnitude limits, $i$ $<$ 19.8 mag and $i$ $<$ 22 mag. For the subset of sources with $W1$ and $W2$ detections, the $W1-W2$ color {\em or} \texttt{XDQSOz} method {\em combined} with variability gives the highest completenesses of $>$ 85\% for both $i$-band magnitude limits and efficiencies of $>$ 80\% to the bright limit and $>$ 60\% to the faint limit; however, the $giW1$ and $giW1$+variability methods give the highest quasar surface densities. The \texttt{XDQSOz} method and combinations of $W1W2$/$giW1$/\texttt{XDQSOz} with variability are among the better selection methods when both high completeness and high efficiency are desired. We also present the OzDES Quasar Catalog of 1,263 spectroscopically-confirmed quasars from three years of OzDES observation in the 30 deg$^{2}$ of the DES supernova fields. The catalog includes quasars with redshifts up to z $\sim$ 4 and brighter than $i$=22 mag, although the catalog is not complete up this magnitude limit.
\end{abstract}

\keywords{dark energy survey, quasar selection}
\maketitle

\section{Introduction}
\label{sec:intro}
\noindent
Quasars are highly energetic sources located at the centers of galaxies created by accretion of matter onto supermassive black holes. They are the most luminous subclass of active galactic nuclei (AGN) and one of the most luminous classes of objects in the Universe. Through accurate measurements of the quasar luminosity function and its evolution, the formation history of supermassive black holes can been studied in detail (\citealp{Kelly2012} and references therein). For studies of baryonic acoustic oscillations (BAO) and the Ly-$\alpha$ forest, quasars act as tracers and backlights of matter clustering (e.g. \citealp{Dawson2013, Ribera2014, Delubac2015}). All these studies benefit from efficient and/or complete selection of large quasar samples. 

\cite{Sandage1965} pioneered quasar selection through the use of multi-color imaging data. Because quasars are not characterized by a single temperature like stars, they usually occupy different regions of color space. For instance, most quasars are bluer in the UV/visible and redder in the infrared. By virtue of this, color selection methods using UV, visible, and mid-infrared photometry are one of the easiest and most common ways to select quasars at various redshifts (e.g. \citealp{Richards2002, Stern2012, Assef2013, Reed2015}). 

Nearly all quasars show $\sim$10\%--20\% stochastic variability at UV and visible wavelengths over timescales of many months to years \citep{KooKron1986, Hook1994, Vries2003, Berk2004,Kelly2009,mcleod2010,Kozlowski2010}. Since only a small fraction of stars are variable at this level, and many of these variable stars are periodic, variability data strongly separates quasars from stars. Models of variability such as a power-law \citep{Schmidt2010,Natalie2011} or a damped random walk \citep{mcleod2010,Butler2011, Ivezic2014} are usually used to identify the quasars. With the emergence of time-domain surveys such as Pan-STARRS \citep{Kaiser2010} and the future Large Synoptic Sky Telescope \citep{Ivezic2008}, quasar selection based on variability will increase in significance and potentially help to fill the selection gaps of color selection techniques. 


In addition to color and variability selection, more sophisticated selection methods have been developed, such as full multi-wavelength SED fitting \citep{Chung2014}, kernel density estimation \citep{Richards2009}, the likelihood method \citep{Kirkpatrick2011like}, neural networks \citep{Yeche2010},  and extreme deconvolution \citep{Bovy2011A}. These statistical methods model the underlying flux  distribution of quasars based on empirical data and then assign probabilities that sources are quasars. The modeled phase space is expandable and can include variability in addition to flux. Statistical quasar selection methods have proven to be very efficient, as they incorporate multi-dimensional information on quasar properties \citep{Ross2012}. 
     
We aim to quantify various quasar selection methods for luminous and point source quasars such as color selection, probabilistic selection, variability selection, and combinations of the selection methods in the Dark Energy Survey (DES) supernova fields. The Dark Energy Survey (DES) is a 5000 deg$^{2}$ \textit{grizY-}band survey of the Southern sky to probe the nature of dark energy \citep{Flaugher2005,Frieman2013}. DES includes ten fields with a total area of $\sim$ 30 deg$^{2}$ that are surveyed at a higher cadence to search for Type Ia supernovae, and are known as the DES supernova fields. We also present a catalog of spectroscopically-confirmed quasars observed by OzDES \citep{Yuan2015} in the DES supernova fields. OzDES is a complementary spectroscopic survey that targets the supernova fields to obtain redshifts for supernovae host galaxies and to conduct quasar reverberation mapping experiment \citep{King2015} and other projects.

The outline of the paper is as follows. In \S\ref{sec:data}, we describe our photometric data sets and their corresponding surveys. These surveys include the Dark Energy Survey (DES), the Vista Hemisphere Survey (VHS; \citealp{McMahon2013}), and WISE \citep{Wright2010}. We introduce the SDSS Stripe 82 quasar catalog from \cite{Peters2015} in \S\ref{sec:peters_cat} that we use to evaluate the quasar selection methods. In \S\ref{sec:shape_selection}, we investigate selecting quasars as point sources. We evaluate the selection completeness and efficiency of various visible and IR color selection methods in \S\ref{sec:color_selection}. We analyze the \texttt{XDQSOz} probabilistic quasar selection algorithm \citep{Bovy2011B, Bovy2012, DiPompeo2015} in \S\ref{sec:xdqso}. In \S\ref{sec:variability}, we consider a $\chi^{2}$-based variability selection method. We present the OzDES survey and the OzDES Quasar Catalog in \S\ref{sec:ozdes_qso_cat}. Finally, we close in \S\ref{sec:discussion} with discussions and conclusions. Throughout the paper, we adopted a flat $\Lambda$CDM cosmology with $H_{0}$ = 70 km s$^{-1}$Mpc$^{-1}$ and $\Omega_{0}$=0.3. Unless stated otherwise, all visible magnitudes refer to the DES magnitudes. The AB magnitude system is used when quoting SDSS and DES magnitudes, while the Vega magnitude system is used for WISE and VHS magnitudes. 

\section{Photometry}
\label{sec:data}
\subsection{Dark Energy Survey}
\label{subsec:des}
The Dark Energy Survey (DES) is a wide-area 5000 deg$^{2}$ survey of the southern hemisphere in the \textit{grizY} bands \citep{Flaugher2005, Frieman2013}. Using the Dark Energy Camera (DECam; \citealp{Flaugher2015}) at the 4m Blanco telescope at the Cerro Tololo Inter-American Observatory, DES aims to probe the nature of dark energy using four different astrophysical probes: Type Ia supernovae, baryonic acoustic oscillations (BAO), galaxy clusters, and weak lensing. The planned 5$\sigma$ point source depths of the survey are \textit{g} = 26.5 mag, \textit{r} = 26 mag, \textit{i} = 25.3 mag, \textit{z} = 24.7 mag, and \textit{Y} = 23 mag \citep{mohr2012}. DES is covering much more area than other surveys of similar depth (e.g. the NOAO Deep Wide-Field Survey \footnote{\url{http://www.noao.edu/noao/noaodeep/}}, \citealp{Jannuzi1999}) and is much deeper than other surveys of larger area (e.g. SDSS and Pan-STARRS), so it is well suited to identifying new quasars. The survey finished its third season of operation in February 2016 \citep{Diehl2016} and recently started its fourth observing season in August 2016. 

DES conducts a multi-epoch supernova survey of two deep fields (C3 and X3) and eight shallow fields to search for Type Ia supernovae \citep{Bernstein2012}. Each supernova field subtends $\sim$ 3 deg$^{2}$, or 30 deg$^{2}$ total. The supernova survey has a mean cadence of $\sim$ 7 days in the \textit{griz} bands. While the main DES survey is conducted under good seeing conditions (at a median FWHM $\sim$ 0\farcs9), it switches to imaging the supernova fields when the seeing increases to $\gtrsim$ 1\farcs1 or when the supernova fields have not been observed for more than a week. The supernova component is expected to comprise roughly one-third or $\sim$ 1300 hours of the total DES observing time. Table \ref{tab:des_sn} shows the names and field centers of the ten supernova fields. Most of the supernova fields are also well-studied by other surveys such as the Chandra Deep Field-South (CDFS; \citealp{Xue2011}) and the VIMOS-VLT Deep Survey (VVDS; \citealp{LF2005}). These fields are optimal for studying variability-based quasar selection since they have more photometric epochs than the main DES wide-field survey. By the end of the 5-year survey, all ten supernova fields are expected to have more than ten times the exposure time of the wide-field survey, more than 100 epochs, and reach an expected 5$\sigma$ point source depth in the \textit{griz} bands of $\sim$ 28 mag for the deep supernova fields and $\sim$ 26.5 mag for the shallow fields \citep{Bernstein2012}. 


For our work, we used the coadded catalog \textit{Y1A1\_COADD\_OBJECTS\_DFULL} for the DES year one (Y1, begun in 2013B) observations of the supernova fields. The catalog combines all available exposures (300 or more) from the DES Y1 operation and some from the Science Verification phase that have sufficient image quality (typical seeing FWHM $<$ 1\farcs1, or 1\farcs25 in some cases). It is typically $\sim$ 2 magnitudes deeper than the DES Y1 wide field coadded catalog. The catalog includes the weighted average of many photometric quantities from single-epoch DES data. For example, \textit{wavgcalib\_mag\_psf} and \textit{wavg\_magerr\_psf} are the weighted average of the single-epoch \textit{mag\_psf} and \textit{magerr\_psf} values. The weighted averages are often found to be more accurate than the coadded quantities derived from SExtractor \citep{Bertin1996}, especially in areas where the number of epochs is small or where the coadded PSFs are not well-fit by PSFEx \citep{Bertin2011}. For the rest of this paper, we used the weighted average quantities whenever referring to DES photometry, such as \textit{wavgcalib\_mag\_psf\_[grizY]} and \textit{wavg\_magerr\_psf\_[grizY]} for the magnitudes and magnitude errors. 

\begin{table} [t]
\small
\label{tab:des_sn}
\caption{DES supernova fields} 
\centering 
\begin{tabular}{c l l l}
\hline\hline 
Field Name & RA & DEC & Depth \\
& h m s & $\arcdeg$ $\arcmin$ $\arcsec$\\
\hline  
E1 & 00 31 29.9 & $-$43 00 34.6 & shallow \\[0.5ex]
E2 & 00 38 00 & $-$43 59 52.8 & shallow\\[0.5ex]
S1 & 02 51 16.8 & $+$00 00 00 & shallow\\[0.5ex]
S2 & 02 44 46.7 & $-$00 59 18.2 & shallow\\[0.5ex]
C1 & 03 37 05.8 & $-$27 06 41.8 & shallow\\[0.5ex]
C2 & 03 37 05.8 & $-$29 05 18.2 & shallow\\[0.5ex]
C3 & 03 30 35.6 & $-$28 06 00 & deep\\[0.5ex]
X1 & 02 17 54.2 & $-$04 55 46.2 & shallow\\[0.5ex]
X2 & 02 22 39.5 & $-$06 24 43.6 & shallow\\[0.5ex]
X3 & 02 25 48.0 & $-$4 36 00 & deep\\[0.5ex]
\hline
\end{tabular}
\end{table}

\subsection{Vista Hemisphere Survey}
\label{subsec:vhs}
The Vista Hemisphere Survey (VHS; \citealp{McMahon2013})\footnote{\url{http://www.vista-vhs.org/}} is a wide-field near-infrared (NIR) survey of the Southern hemisphere to a depth 30 times deeper than 2MASS \citep{Kleinmann1994} in the \textit{J} and \textit{K} bands. Some parts of the sky are also imaged in the \textit{Y} and \textit{H} bands. VHS aims to cover 18,000 deg$^{2}$ of the Southern hemisphere and overlaps $\sim$ 4500 deg$^{2}$ of the DES footprint in the South Galactic Cap. It is the deepest near infrared survey that overlaps a large fraction of the DES footprint. In the overlap region, VHS has a median 5$\sigma$ point source detection depth of \textit{J$_{\rm Vega}$} = 20.3 mag and \textit{K$_{\rm Vega}$} = 18.6 mag with 80\% completeness \citep{Banerji2015}. We used the publicly available Data Release 3 source/merged catalog\footnote{\url{http://horus.roe.ac.uk/vsa/}} \textit{vhsSource} and the default point source aperture corrected magnitude and error, \textit{xAperMag3} and \textit{xAperMag3Err} where \textit{x} refers to the desired band.

\subsection{WISE}
\label{subsec:wise}
The Wide-Field Infrared Survey Explorer (WISE; \citealp{Wright2010}) is an all-sky survey of the mid-IR sky at 3.4, 4.6, 12, and 22 $\micron$ (W1, W2, W3, and W4). After it depleted its hydrogen cryogen in 2010, the WISE mission was renamed NEOWISE \citep{Mainzer2011} and it continued to survey the sky in the W1 and W2 bands until 2011. It was then reactivated in 2013 to conduct a ``post-cryogenic'' three-year survey of the sky in the W1 and W2 bands \citep{Mainzer2014}. At present, over 99\% of the sky has $\geq$ 23 exposures in these two bands (Myers et al. 2015).

We used the instrumental profile-fit photometry from the \textit{AllWISE} source catalog\footnote{\url{http://wise2.ipac.caltech.edu/docs/release/allwise/}}\footnote{\url{http://irsa.ipac.caltech.edu/cgi-bin/Gator/nph-dd}}, i.e. \textit{wxmpro} (magnitude), \textit{wxsigmpro} (magnitude error), and \textit{wxsnr} (signal-to-noise ratio) where \textit{x} is 1 or 2, for the rest of this paper. The \textit{AllWISE} catalog combines data from both the WISE cryogenic and the NEOWISE post-cryogenic phase, which improves the W1 and W2 sensitivities. Magnitude limits for the \textit{ALLWISE} catalog are 16.9 mag and 16.0 mag in W1 and W2, respectively, in the Vega system \citep{Wright2010,Stern2012}.

\section{Peters et al. (2015) Quasar Catalog}
\label{sec:peters_cat}
We used a subset of the SDSS Stripe 82 quasar catalog from \cite{Peters2015} (hereafter referred to as P15) that overlaps with the DES supernova fields for our analyses of the quasar selection methods. P15 used a Bayesian analysis combining color and variability to identify 36,569 quasar candidates in the SDSS Stripe 82 field, of which 35,820 (98\%) are considered to be ``good'' quality candidates that pass their color cuts for removing stellar and white dwarf contaminants. Of these ``good'' candidates, 36\% are spectroscopically-confirmed, and 92\% of the spectroscopically-confirmed quasars are brighter than coadded $i_{\rm SDSS}$ = 19.9 mag. Using that sample, P15 estimated a (spectroscopic) completeness of 94.3\% for the overall ``good'' quasar candidate sample at $i_{\rm SDSS}$ = 19.9 mag (in agreement with \citealp{Berk2005}). The rest of the ``good'' candidates with no spectroscopic confirmation reach a depth of \textit{i$_{\rm SDSS}$} $\sim$ 22 mag; their redshifts are estimated using optical photometry and astrometry. P15 have limited spectroscopic data to measure the completeness and efficiency of their ``good'' candidates, including deeper spectroscopy for $\sim$ 5000 quasars with 19 $<$ $i_{\rm SDSS}$ $<$ 22 mag from the BOSS DR10 and DR12 quasar catalogs. The P15 ``good'' candidates recover $\sim$97\%--98\% of these spectroscopically confirmed quasars, so they conclude that their sample is $\sim$ 97\% complete up to $i_{\rm SDSS}$ $<$ 22 mag. We conduct our quasar selection analyses using both the bright (\textit{i$_{\rm SDSS}$} $<$ 19.9 mag) and total ($i_{\rm SDSS}$ $\lesssim$ 22 mag) samples.


SDSS Stripe 82 has a 5.7 deg$^{2}$ overlap with the DES supernova fields in the S1 and S2 fields. For our selection analyses, we only considered the P15 catalog within this region of overlap and call this subset the P15-S1S2 Quasar Catalog. Within the overlap region, the P15-S1S2 Quasar Catalog contains 975 quasar candidates. We cross-matched the P15-S1S2 Quasar Catalog with the DES \textit{Y1A1\_COADD\_OBJECTS\_DFULL} catalog (see \S\ref{subsec:des}) with a 0\farcs5 matching radius, resulting in 900 matches (a larger matching radius of 1\farcs0 returns two extra matches). From the 900 matches, 671 (120) have $i$ $<$ 22 mag ($i$ $<$ 19.8 mag), of which 60\% (95\%) have spectroscopic redshifts from the P15 catalog. Using the color terms in \S\ref{sec:variability}, we estimated \textit{i$_{\rm SDSS}$} = 19.9 mag to be $i$ $\sim$ 19.8 mag. 

\begin{figure*}[tb]
	\centering
    \includegraphics[width=0.9\textwidth]{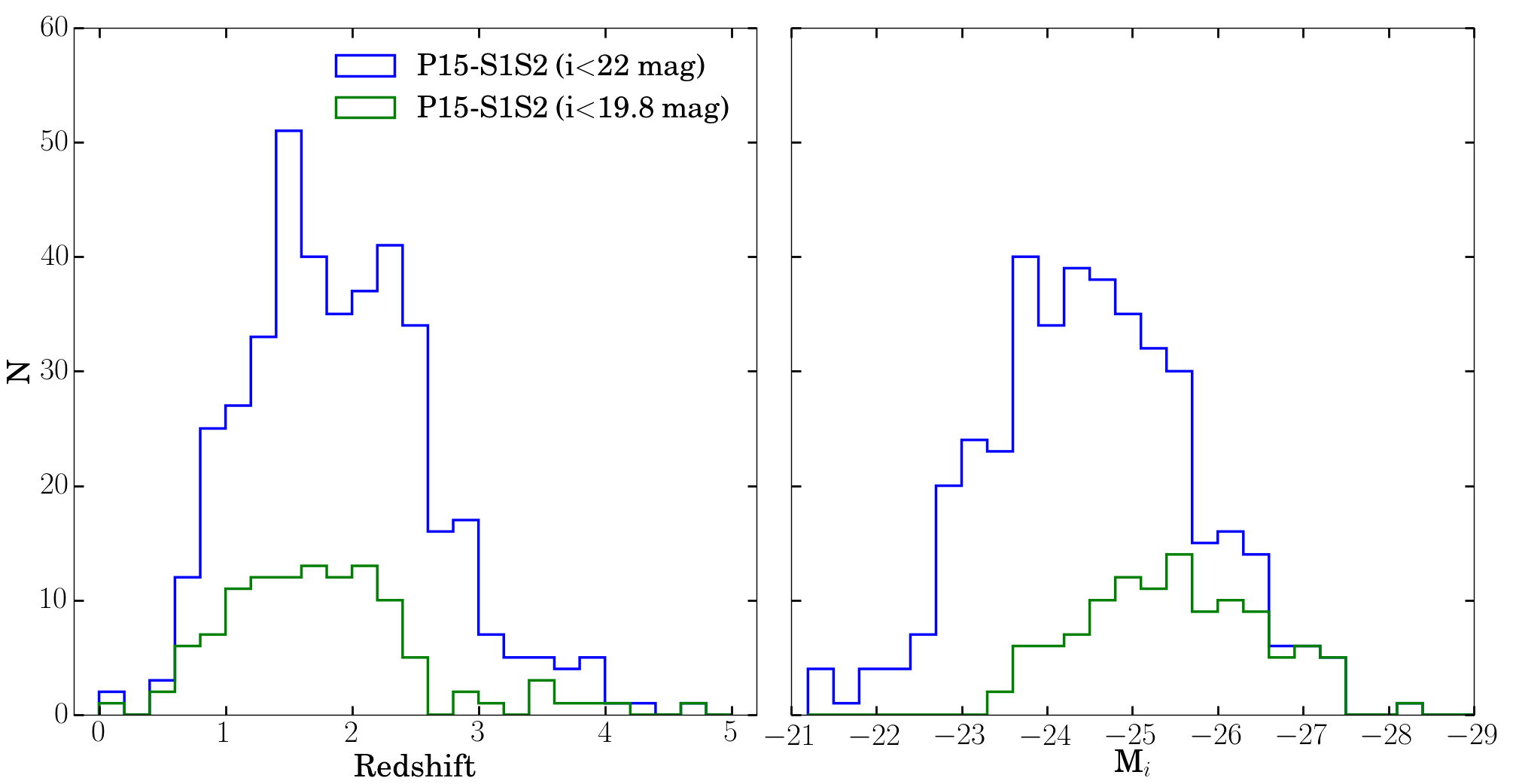}
    \label{fig:peters_Miz}
    \caption{Distribution of spectroscopic redshifts and absolute $i$-band magnitudes for the bright and total samples of the P15-S1S2 Quasar Catalog with DES matches. The absolute magnitudes are calculated using DES $i$-band magnitudes and k-corrected spectroscopic redshifts from the left panel, with k-corrections from \cite{Richards2006A}.}
\end{figure*}

\begin{figure*}[tb]
	\centering
    \includegraphics[width=0.9\textwidth]{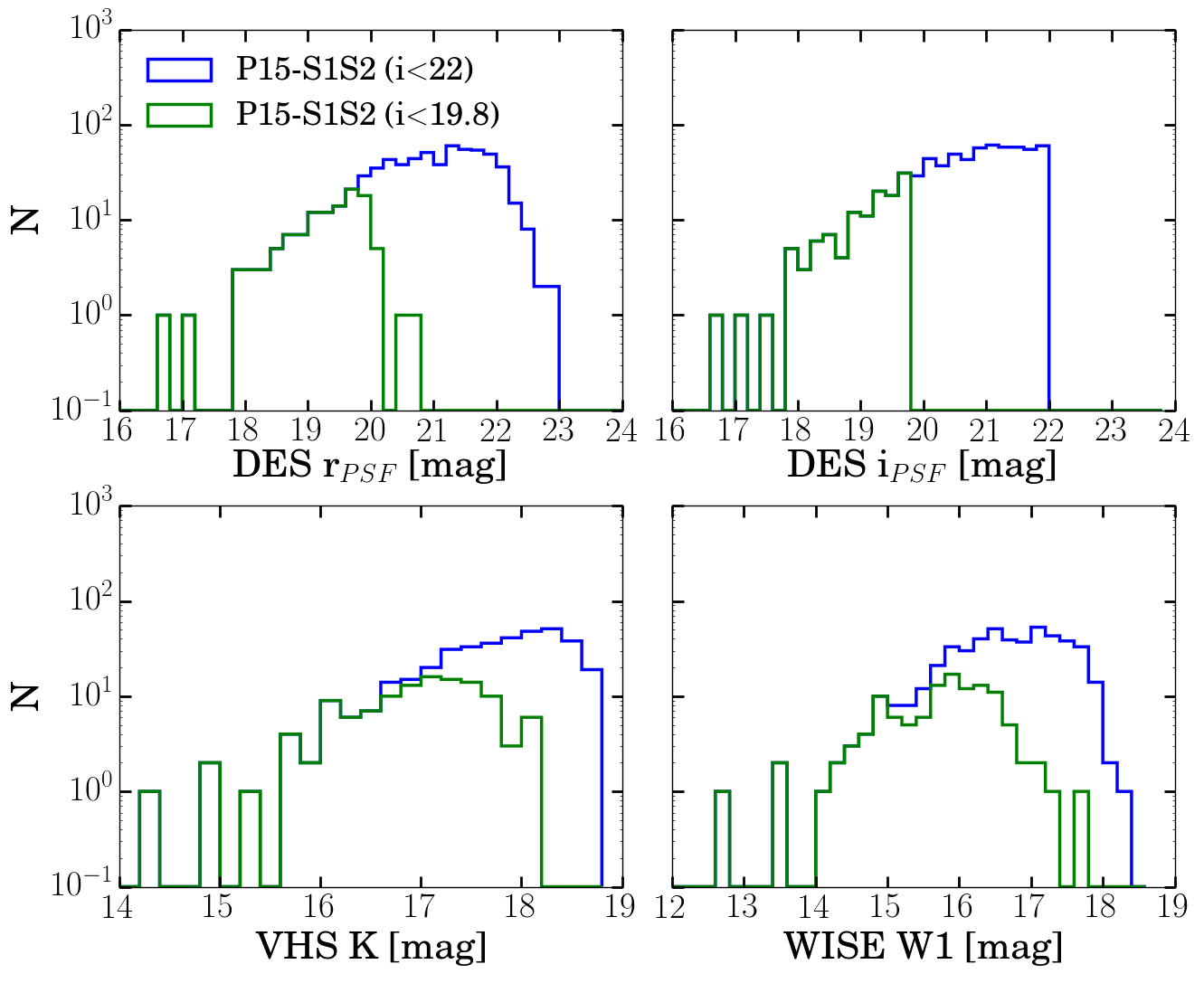}
    \label{fig:peters_mag}
    \caption{Distribution of the P15-S1S2 quasars with DES matches in the visible to the mid-IR wavelengths.}
\end{figure*}

We treated the P15-S1S2 catalog as a truth catalog for analyses of our selection methods at the bright ($i$ $<$ 19.8 mag) and faint ($i$ $<$ 22 mag) limits. We also defined a corresponding catalog of non-quasars in the overlapping DES S1 and S2 fields at both magnitude limits, which consists of point sources that are not in the P15-S1S2 Quasar Catalog. We define point sources in \S\ref{sec:shape_selection} and use the point source cut to reduce contamination from extended sources. We cross-matched the P15-S1S2 catalog and the non-quasar catalog with VHS and WISE with matching radii of 0\farcs5 and 3\farcs0, respectively (a larger matching radius of 1\farcs0 with VHS returns $\sim$ 1\% extra matches and $\sim$ 80\% of the matches with WISE are within 1\farcs0). Figure \ref{fig:peters_Miz} shows the spectroscopic redshift and absolute $i$-band magnitude distributions for the matched total and bright samples of the P15-S1S2 Quasar Catalog while Figure \ref{fig:peters_mag} shows their DES, WISE, and VHS magnitude distributions.

\begin{figure}[tb]
	\centering
    \includegraphics[width=0.45\textwidth]{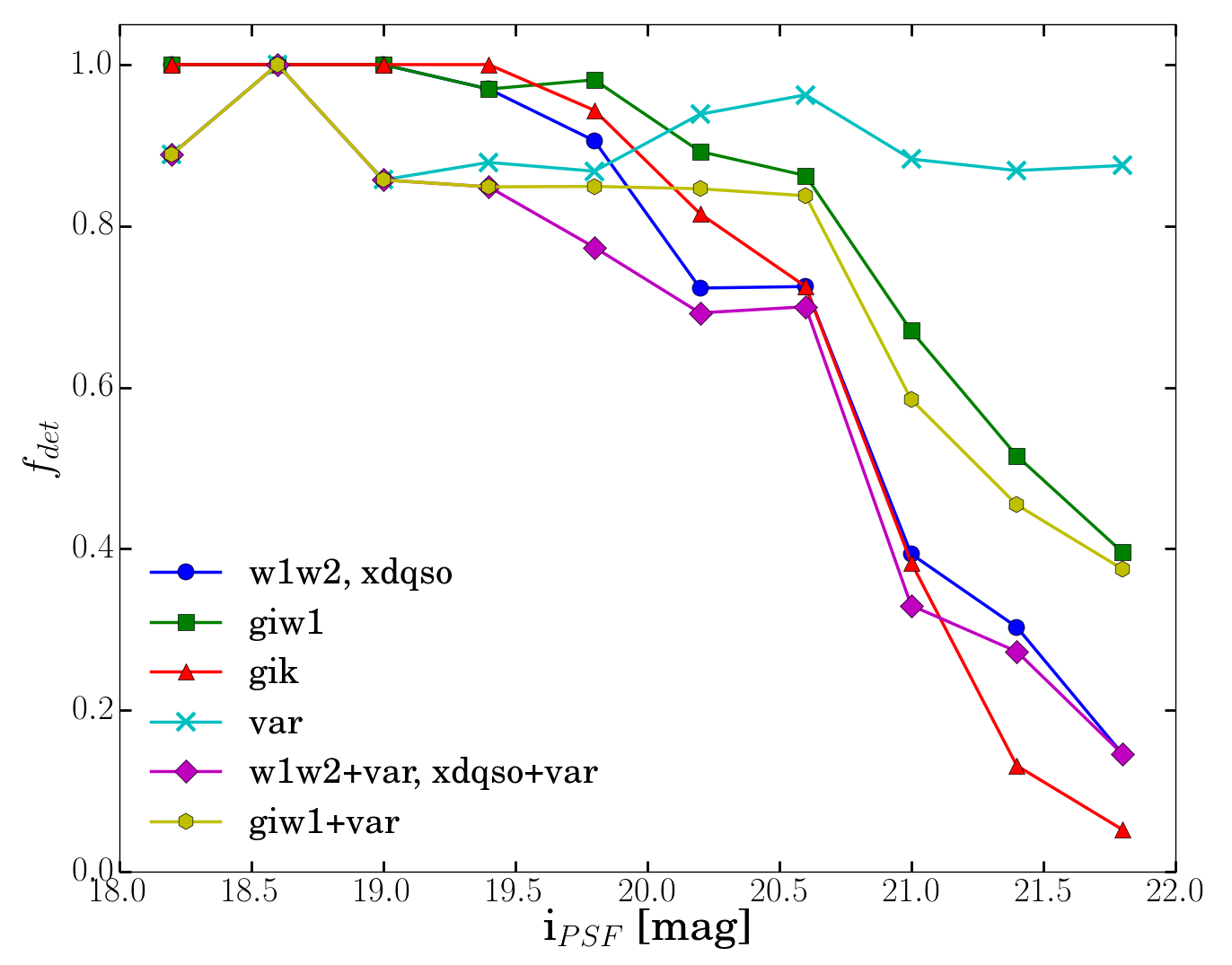}
    \label{fig:normalization}
    \caption{The fraction $f_{det}$ of P15-S1S2 quasars that meet the selection-dependent detection threshold(s) relative to the quasars that satisfy the flag, magnitude error, and point source cuts. The lines are color-coded according to the type of selection and illustrate the consequences of requiring WISE, VHS, or epoch limits on the sample size. The $W1W2$ and \texttt{XDQSO} methods have the same $f_{det}$ curve as they both require detections in the $W1$ and $W2$ bands.}
\end{figure}

We imposed a set of preliminary cuts to define the underlying sample from which to calculate the completeness and efficiency of the selection methods in this work. Objects are required to be point sources, have DES $flags\_[griz]$ $<$ 3 in at least one of the \textit{griz} bands (i.e. not close to a bright neighbor and not originally blended), have magnitude errors $<$ 1 mag in the $gri$ bands (i.e. detected in those bands), and detected in all the required non-DES selection bands. 
Figure \ref{fig:normalization} shows the fraction of the P15-S1S2 quasars that satisfy the detection cut(s) for each selection method as a function of $i$-band magnitude. There are 671 total (120 bright) P15-S1S2 quasars, of which 563 (105) satisfy the point source, flag and magnitude error cuts, where the flag cut removes $\sim$ 11\% of the $i$ $<$ 22 mag and 8\% of the $i$ $<$ 19.8 mag quasars. Of these 563 (105) quasars, 405 (103) have WISE $W1$ detections, 308 (101) have WISE $W1$ and $W2$ detections, and 291 (104) have VHS $K$-band detections. Note that the median $K$-band depth in the P15-S1S2 region is shallower than the median VHS depth over the wider area quoted in \cite{Banerji2015}. 

We give the completeness and efficiency for all the selection methods relative to sources that pass the preliminary cuts. Completeness is defined as the fraction of P15-S1S2 quasars that are selected, for the subset of the P15-S1S2 quasars that have sufficient data to apply the selection method. In other words, the numerator and denominator are only comprised of objects with sufficient data to apply the selection method. Efficiency is defined as the number of selected quasars divided by the total number of sources that satisfied the selection criterion. We will report the results for the selection methods for the full \textit{i} $<$ 22 mag sample followed by the results for the bright sample \textit{i} $<$ 19.8 mag in parenthesis.

\section{Selection of Point Sources}
\label{sec:shape_selection}
Due to the point-like appearance of quasars in most ground-based imaging data, shape information can be used to reduce host galaxy contamination. We used the DES weighted average of the SExtractor \citep{Bertin1996} \textit{spread\_model} star/galaxy classifier, which is a discriminant between the best-fitting PSF model and a more extended model \citep{Desai2012}. Specifically, we used the cut \textbar \textit{wavg\_spread\_model\_r}\textbar $<$ \textit{0.003 + spreaderr\_model\_r} to identify point sources \citep{Drlica2015}. 

We calculated the completeness of the shape selection criterion using a sample of very high confidence point sources from regions of the Canada-France-Hawaii Telescope Lensing Survey (CFHTLenS; \citealp{Heymans2012, Erben2013}) that overlap three of the DES supernova fields (X1, X2, and X3). CFHTLenS combines five years worth of data from the CFHT Legacy Survey and has reliable shape estimates calculated as part of the optimization for weak lensing analyses.  We used the \textit{class\_star} and \textit{lensfit} \citep{Miller2012} parameters from the public CFHTLenS catalog to select point sources, defined to be sources with \textit{class\_star} $>$ 0.98 or \textit{lensfit} = 1, and analyzed the three supernova fields separately, as the fields have different depths with the X3 field being the deepest. We tested our shape selection on the P15-S1S2 Quasar Catalog. All CFHTLenS objects and P15-S1S2 quasars are required to have DES $flags\_[griz]$ $<$ 3 in at least one of \textit{griz} bands to ensure the sources are not close to bright neighbors and not originally blended. Additionally, the sources need to have DES magnitude errors of $<$ 1 mag in the \textit{gri} bands, where a magnitude error of 1 mag corresponds to non-detection. 

\begin{figure*}[tb]
	\centering
    \includegraphics[width=0.9\textwidth]{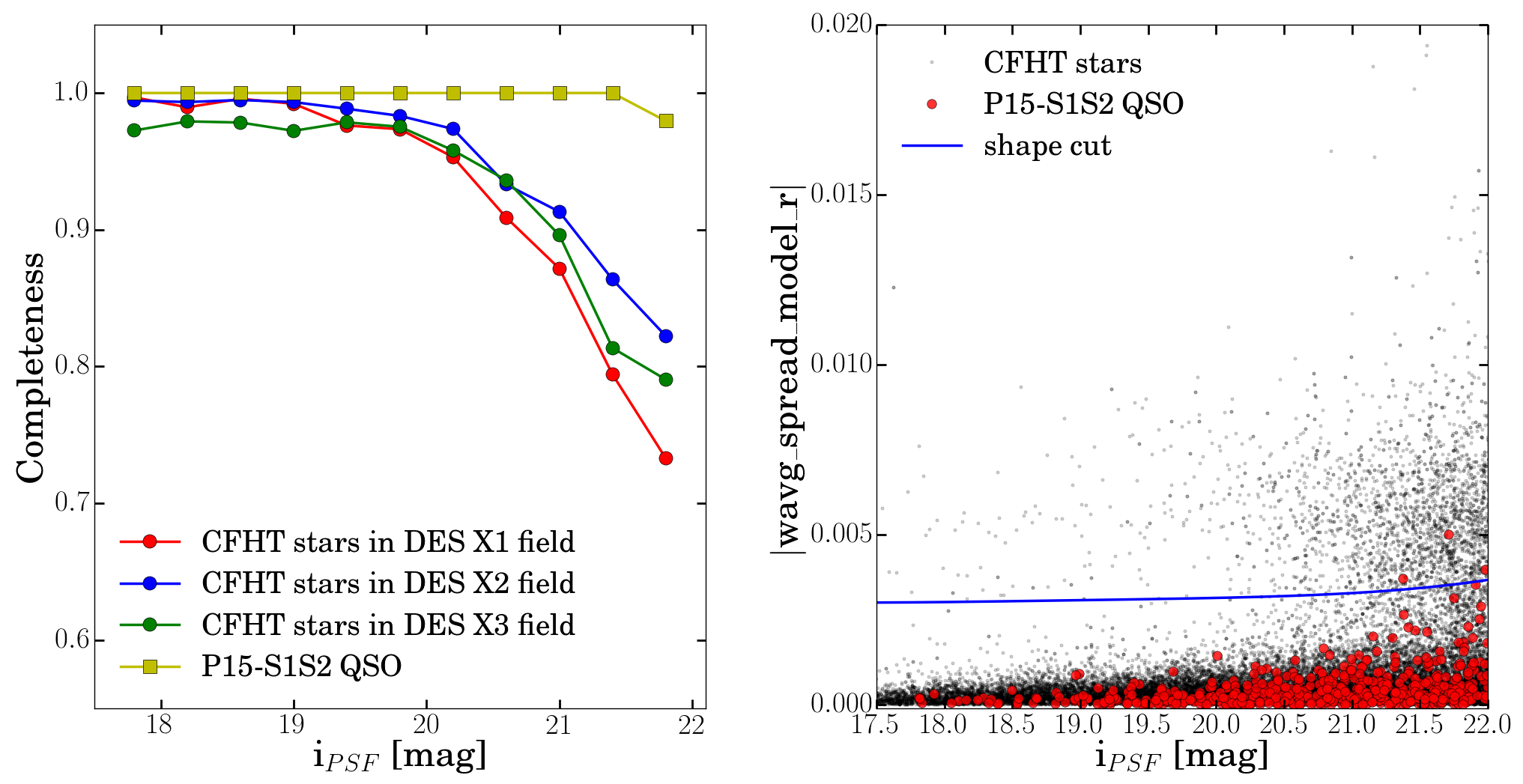}
    \label{fig:shape}
    \caption{\textit{Left panel:} Fraction of objects classified as stars by CFHTLenS that are also classified as point sources by the cut \textbar \textit{wavg\_spread\_model\_r}\textbar $<$ \textit{0.003 + spreaderr\_model\_r} as a function of DES $i$-band magnitude. Most of the P15-S1S2 quasars satisfy this shape cut, with only 0.4\% failing at $i$ $<$ 22 mag, reaching a completeness of 99.6\%. \textit{Right panel:} Shape information for the CFHTLenS point sources and quasars from the P15-S1S2 Quasar Catalog as a function of $i$-band magnitude. Gray points are the CFHTLenS point sources while red points are the P15-S1S2 quasars. The blue line is a representation of the shape cut that illustrates how it becomes more lenient at fainter magnitudes due to the increase in \textit{spreaderr\_model\_r}. }
\end{figure*}

The left panel of Figure \ref{fig:shape} shows the fraction of CFHTLenS point sources and P15-S1S2 quasars that pass our point source selection criterion as a function of $i$-band magnitude in the three DES supernova X-fields. For selecting CFHTLenS stars, the completeness exceeds 90\% for \textit{i} $\lesssim$ 20.5 mag and then drops to $\sim$ 80\% by \textit{i}=22 mag. At \textit{i} $>$ 20 mag, the stellar sources from CFHTLenS are likely incomplete and the uncertainties in \textit{spreaderr\_model\_r} uncertainties are becoming large. The right panel of Figure \ref{fig:shape} shows the \textit{wavg\_spread\_model\_r} shape information for the CFHTLenS stars and P15-S1S2 quasars as a function of $i$-band magnitude. The blue line illustrates the typical behavior of the shape cut as a function of magnitude using the median value for CFHTLenS stars in magnitude bins of $\Delta m = 0.4$ mag. It becomes more forgiving at fainter magnitudes due to the increase in \textit{spreaderr\_model\_r}. The exact cut varies even for objects with the same apparent magnitude due to variations in the data such as the background and the number of coadd tiles. The shape cut selects quasars from the P15-S1S2 Quasar Catalog with a completeness of 99.6\% for \textit{i} $<$ 22 mag. We apply the shape cut in the selection analyses to select point sources. 


\section{Color Selection}
\label{sec:color_selection}
Quasars have a non-thermal continuum with UV/optical spectra that can be approximated by a power-law, in contrast to the single-temperature blackbody that roughly describes most stars. As a result, quasars tend to occupy distinct regions of color-color space compared to stellar sources \citep{Richards2002, Croom2004, Lacy2004, Maddox2012, Stern2012, Assef2013}. Color cuts are thus commonly used to select quasars. Low-redshift (z $\lesssim$ 2) quasars are frequently selected using \textit{u}-band data that identify the bright ultraviolet continuum emission of quasars. At z $>$ 2.2, the Ly$\alpha$ line exits the $u$-band and enters the bluer $b$ or $g$ optical bands, resulting in redder $u -b/g$ colors which are similar to the color of stars. Although the wide-field region of DES lacks \textit{u}-band, many of the most interesting problems require finding quasars at higher redshifts where this selection approach fails. We focus on combinations of optical, near-IR, and mid-IR data that can effectively identify both high and low redshift quasars even in the absence of $u$-band data. 

We evaluate three color selection methods $W1-W2$, $g-i$ vs. $i-W1$, and $g-i$ vs. $i-K$ using the P15-S1S2 Quasar catalog. The $giW1$ and $giK$ color selections were initially used or developed to target AGNs for the OzDES reverberation mapping monitoring campaign and they were designed based on the stellar locus to reduce contamination from stars.  

\subsection{WISE $W1-W2$ color section}
\label{subsec:w1w2_selection}
Quasars generally have a red $W1-W2$ color due to emission from hot dust (in the case of low-redshift quasars) or the accretion disk (in the case of high-redshift quasars), unless a strong emission line happens to enter the $W1$ band \citep{Assef2010}. Contaminants with similarly red $W1-W2$ colors are cool brown dwarfs and dusty stars. For example, brown dwarfs with spectral class cooler than T1 have red colors in these WISE bands due to methane absorption \citep{Cushing2011,Kirkpatrick2011bd}. We investigated quasar selection using $W1-W2$ $>$ 0.7 (Vega magnitudes) based on \cite{Stern2012}. The sample is limited to point sources with a signal-to-noise ratio of SNR $>$ 5 in both the $W1$ ($\sim$ 16.9 mag) and $W2$ ($\sim$ 16 mag) bands.

\begin{figure*}[tb]
	\centering
    \includegraphics[width=0.9\textwidth]{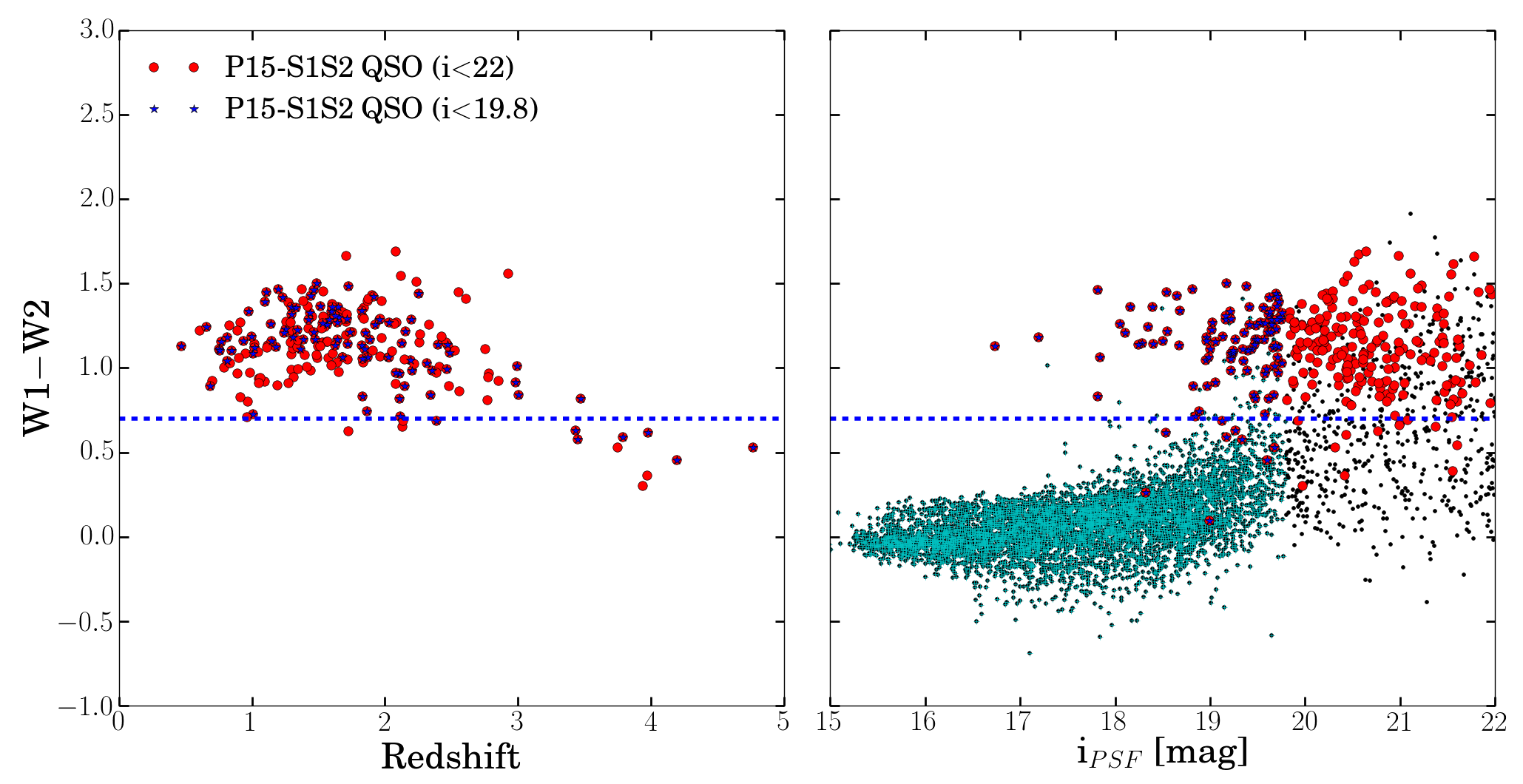}
    \label{fig:w1w2}
    \caption{\textit{Left panel:} WISE $W1-W2$ color as a function redshift for quasars in the P15-S1S2 quasars. \textit{Right panel:} WISE color as a function of the DES $i$-band magnitude for the  same quasars plus point source non-quasars in the DES S1 and S2 fields, where black (cyan) points have \textit{i} $<$ 22 mag ($i <$ 19.8 mag). The WISE color cut is shown by the dotted blue line in both panels. The WISE color is on the Vega system and the $i$-band magnitude is on the AB system. }
\end{figure*}

We calculate the completeness and efficiency of this color selection method relative to the total and bright samples of the P15-S1S2 Quasar Catalog. In each case we give the results for the \textit{i} $<$ 22 mag sample followed by the results for the \textit{i} $<$ 19.8 mag sample in parenthesis. This color selection identifies 287 (92) of the 308 (101) P15-S1S2 quasars with the data necessary to apply this cut, for a completeness of 93\% (91\%). There are also 246 (37) other point sources that satisfy this color selection, for an efficiency of 54\% (71\%). Figure \ref{fig:w1w2} shows the $W1-W2$ colors as a function of redshift and apparent \textit{i}-band magnitude for the P15-S1S2 quasars and the point source non-quasars in the DES S1 and S2 fields. 


\subsection{$giW1$ selection}
\label{subsec:giw1_selection}
\begin{figure}[tb]
\centering
\includegraphics[width=0.5\textwidth]{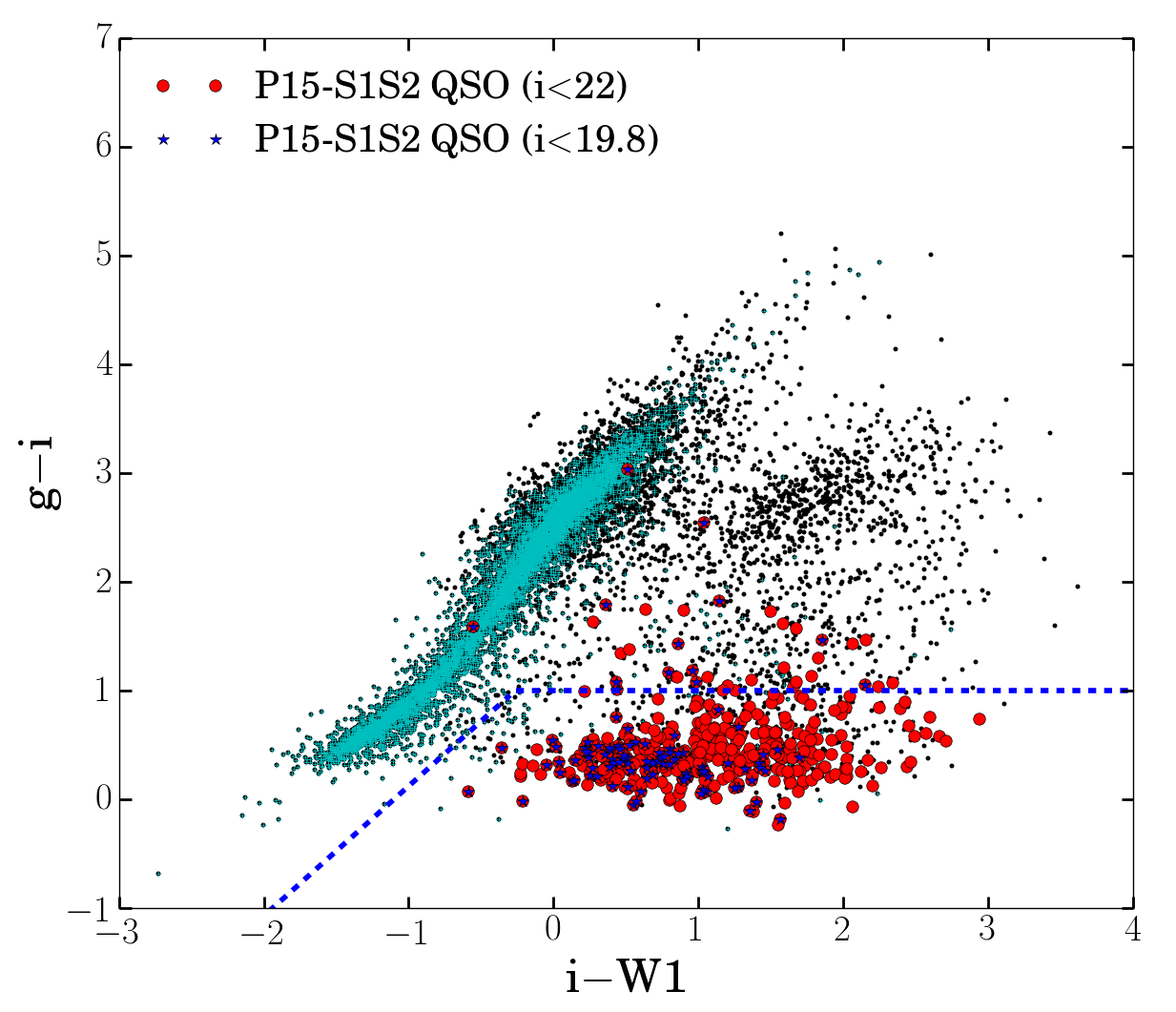}
\label{fig:giw1}
\caption{Color-color diagram of quasars from the P15-S1S2 catalog and point source non-quasars in the DES S1 and S2 fields, where black (cyan) points are point sources with \textit{i} $<$ 22 mag (\textit{i} $<$ 19.8 mag). The blue dotted lines define the selection region. Note that the visible and WISE magnitudes are both in the AB system. }
\end{figure}

Quasars are expected to be blue in the $g$-band compared to redder optical filters, but red when comparing to the mid-IR \citep{Wu2012,Chehade2016}. We therefore investigated a color selection method combining
\begin{gather}
(g - i) < 1 \mbox{ with}\nonumber \\
(g - i) < 1.195*(i - W1_{\rm AB}) + 1.317
\end{gather}

\noindent 
where all the magnitudes are on the AB system, with $W1_{AB} = W1_{Vega} + 2.699$ \footnote{\url{http://wise2.ipac.caltech.edu/docs/release/allsky/expsup/sec4_4h.html\#conv2ab}}. The $g - i$ limit is intended to reduce contamination from compact galaxies, but excludes high-redshift quasars with z $>$ 3.5. This color selection identifies 369 (90) of the 405 (103) P15-S1S2 quasars with the data necessary to apply this cut, for a completeness of 91\% (87\%). There are 391 (58) other point sources that satisfy this color selection, for an efficiency of 49\% (61\%). Figure \ref{fig:giw1} shows the $i-W1$ and $g-i$ colors for quasars in the P15-S1S2 catalog and the point source non-quasars in the S1 and S2 fields. 


\subsection{$giK$ selection}
\label{subsec:gik_selection}
\begin{figure}[tb]
	\centering
	\includegraphics[width=0.5\textwidth]{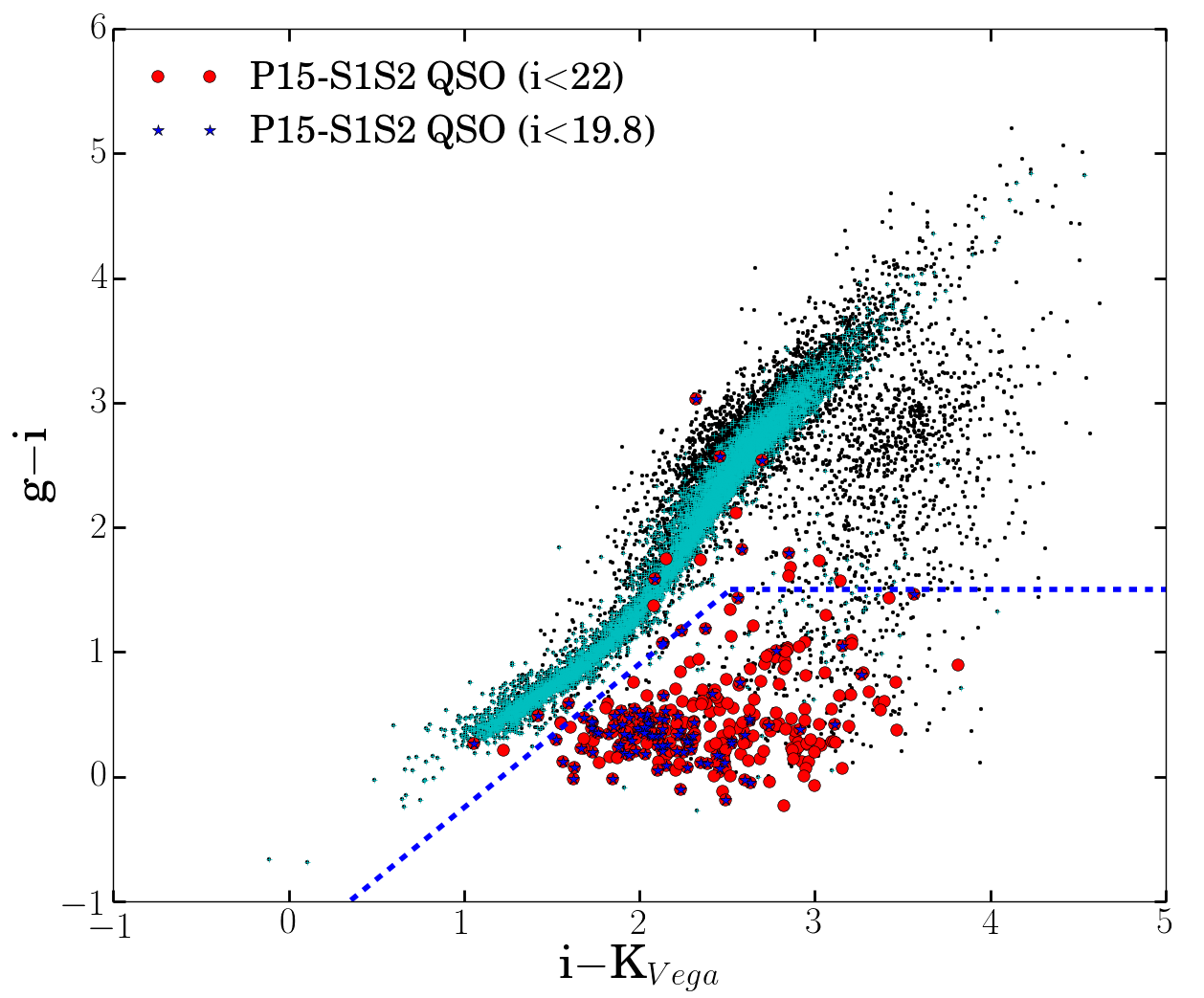}
    \label{fig:giK}
	\caption{Color-color diagram of quasars from the P15-S1S2 catalog and point source non-quasars in the DES S1 and S2 fields. Black (cyan) points are point sources with \textit{i} $<$ 22 mag (\textit{i} $<$ 19.8 mag). The blue dotted lines define the color selection region. The visible magnitudes are on the AB system and the VHS magnitude is on the Vega system. }
\end{figure}

Quasars between 2 $<$ z $<$ 3 have visible-wavelength colors similar to stars, making it challenging to select quasars in this redshift range with high efficiency using color selection. The traditional method to separate such intermediate-redshift quasars from stars is the UV excess method \citep{Richards2002}, as quasars at these redshifts have more UV flux than stars due to the presence of the Ly${\alpha}$ emission line in the UV filters. Here, an alternative approach is needed as DES lacks $u$-band data. Intermediate-redshift quasars also have a near-infrared \textit{K}-band excess compared to stars \citep{Warren2000}, so we tested the color cuts
\begin{gather}
(g - i) < 1.5 \mbox{ with}\nonumber \\
(g - i) < 1.152*(i - K_{\rm Vega}) - 1.4, \mbox{ with \textit{K$_{\rm Vega}$} $<$ 18.6 \mbox{mag}}
\end{gather}
\noindent
following \cite{Banerji2015}. Note that this mixes DES AB with VHS Vega magnitudes. Similarly as \S\ref{subsec:giw1_selection}, the $g - i$  limit which is more relaxed here is used to minimize contamination from compact galaxies but also excludes high-redshift quasars (z $>$ 3.5). The median $K$-band depth in the P15-S1S2 region is also shallower than the median VHS depth over the wider DES-VHS area quoted in \cite{Banerji2015}. This color selection identifies 269 (93) of the 291 (104) P15-S1S2 quasars with the data necessary to apply this cut, for a completeness of 92\% (89\%). There are 344 (55) other point sources that satisfy this color selection, for an efficiency of 44\% (63\%). Figure \ref{fig:giK} shows the $i-K_{\rm Vega}$ and $g-i$ colors for the P15-S1S2 quasars in the P15-S1S2 and other point source non-quasars in the S1 and S2 fields. 


\section{XDQSO Selection}
\label{sec:xdqso}
The change in the observed colors of quasars with redshift restricts the effectiveness of any one set of colors to select quasars to a limited redshift range. Probabilistic techniques have been devised to select quasars more effectively over a broader range of redshifts based on empirical models of quasar and stellar photometry. One such technique is \texttt{XDQSO} \citep{Bovy2012}, which uses density estimation in flux space to assign a quasar probability. \texttt{XDQSO} was developed with a stellar training set from the SDSS Stripe-82 \citep{Abazajian2009} and a quasar training set from the SDSS DR 7 quasar catalog \citep{Schneider2010}. \cite{Bovy2011B} applied \texttt{XDQSO} to the SDSS Data Release 8 \citep{sdssdr8} to create an input quasar catalog for the \textit{BOSS} survey \citep{Ross2012}. They found that \texttt{XDQSO} performs as well as color-based quasar-selection methods at low-redshift (z $<$ 2.2), and better than all other color-based quasar selection methods at intermediate redshifts (2.2 $\leq$ z $\leq$ 3.5). 

\begin{figure*}[tb]
	\centering
    \includegraphics[width=0.9\textwidth]{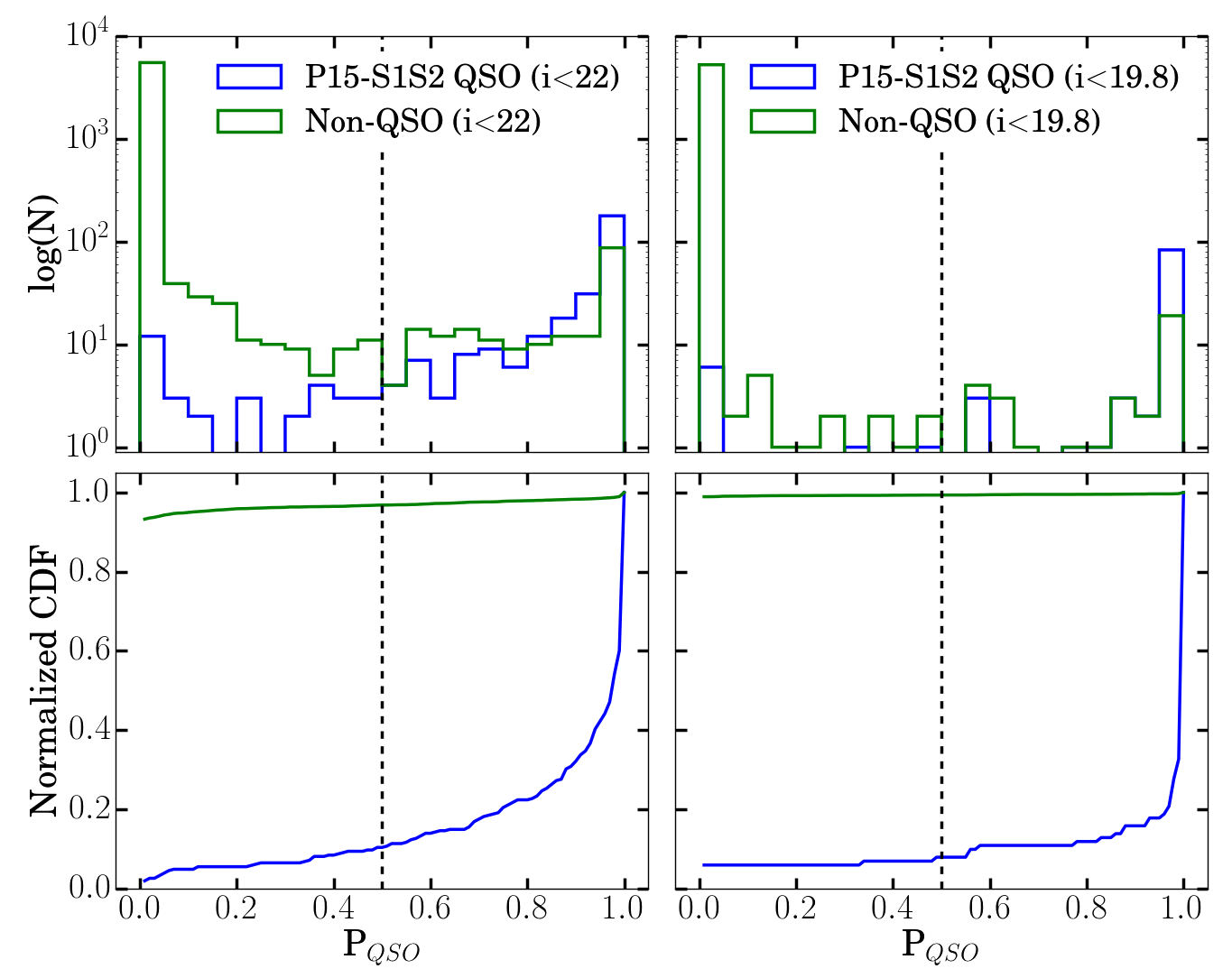}
    \label{fig:xdqso_peters_total}
	\caption{\textit{Top panel:} Distribution of \texttt{XDQSOz} quasar probabilities for the P15-S1S2 quasars and point source non-quasars with WISE photometry at the faint and bright magnitude limits. The dotted line denotes the probability cut at P$_{\rm QSO}$=0.5. \textit{Bottom panel:} The cumulative distribution of the same quasars and non-quasars in P$_{\rm QSO}$.  The majority ($>$ 90\%) of the non-quasars have P$_{\rm QSO}$ $<$ 0.5 while quasars mostly have P$_{\rm QSO}$ $>$ 0.5. }
\end{figure*}

We used the more recent \texttt{XDQSOz} implementation \citep{Bovy2012} combining the DES \textit{griz} magnitudes corrected for Galactic extinction \citep{SF2011} and WISE photometry. We transformed the DES magnitudes to the SDSS systems using color corrections of
\begin{gather}
g_{\rm DES}-g_{\rm SDSS} = -0.083\mbox{ }(g-r)_{\rm DES} - 0.024, \nonumber \\
r_{\rm DES}-r_{\rm SDSS} = -0.083\mbox{ }(g-r)_{\rm DES} - 0.004, \nonumber \\
i_{\rm DES}-i_{\rm SDSS} = -0.352\mbox{ }(i-z)_{\rm DES} + 0.017, \mbox{ and}\nonumber \\
z_{\rm DES}-z_{\rm SDSS} = -0.104\mbox{ }(i-z)_{\rm DES} - 0.007
\end{gather}
\noindent derived from bright SDSS stars\footnote{We used objects from  SDSS \textit{PhotoPrimary} table that are between 17 and 19 magnitudes, have errors less than 0.5 magnitudes, are located in DES S1 and S2 fields, and have SDSS flags satisfying \texttt{!deblend\_too\_many\_peaks \&\& !moved \&\& binned1 \&\& !satur\_center \&\& !bad\_counts\_error \&\& !notchecked\_center \&\& !edge \&\& psf\_flux\_interp}}. Since \texttt{XDQSOz} expects a \textit{u}-band measurement, we supplied a very small flux and an inverse variance of 10$^{-10}$ for this band. We found that including the near-IR data from VHS leads to a slight improvement, but at the expense of a much smaller sample because many sources lack the necessary VHS data. Figures \ref{fig:xdqso_peters_total} shows the \texttt{XDQSOz} probability distributions for the P15-S1S2 quasars and point source non-quasars in the total and bright samples. The \texttt{XDQSOz} selection with a probability cut at P$_{\rm QSO}$ = 0.5 identifies 276 (93) of the 308 (101) P15-S1S2 quasars with the data necessary to apply this cut, for a completeness of 90\% (92\%). There are 185 (34) other point sources that satisfy this color selection, for an efficiency of 60\% (73\%).

\section{Variability Selection}
\label{sec:variability}

We obtained light curves from the DES Y1 single-epoch catalogs, \textit{Y1A1\_IMAGE} and \textit{Y1A1\_FINALCUT}, which span less than a year and typically have $\sim$ 15 epochs. Due to the relatively limited temporal extent of this data, we cannot adopt sophisticated variability models based on fitting structure function models. Instead, we used a multi-band \textit{griz} $\chi^{2}$ variability statistic to distinguish between variable and non-variable sources. We calculated $\chi^{2} = \sum_{K=g,r,i,z}\sum_{j=1}^{N_{K}}(m_{K,j}-\langle m_{K} \rangle)^{2}/\sigma_{K,j}^{2}$ for each source and evaluated the null hypothesis that the source has constant magnitude $\langle m_{K} \rangle$ over $N_{K}$ epochs, where $\langle m_{K} \rangle$ is defined as the error-weighted mean magnitude $\langle m_{K} \rangle = \left(\sum  m_{K,j}/\sigma_{K,j}^{2}\right)/\left(\sum1/\sigma_{K,j}^{2}\right)$ for each band $K$. We imposed a minimum error floor of $\sigma_{j}$ = 0.01 mag for the measurements. 

\begin{figure*}[tb]
	\centering
    \includegraphics[width=0.9\textwidth]{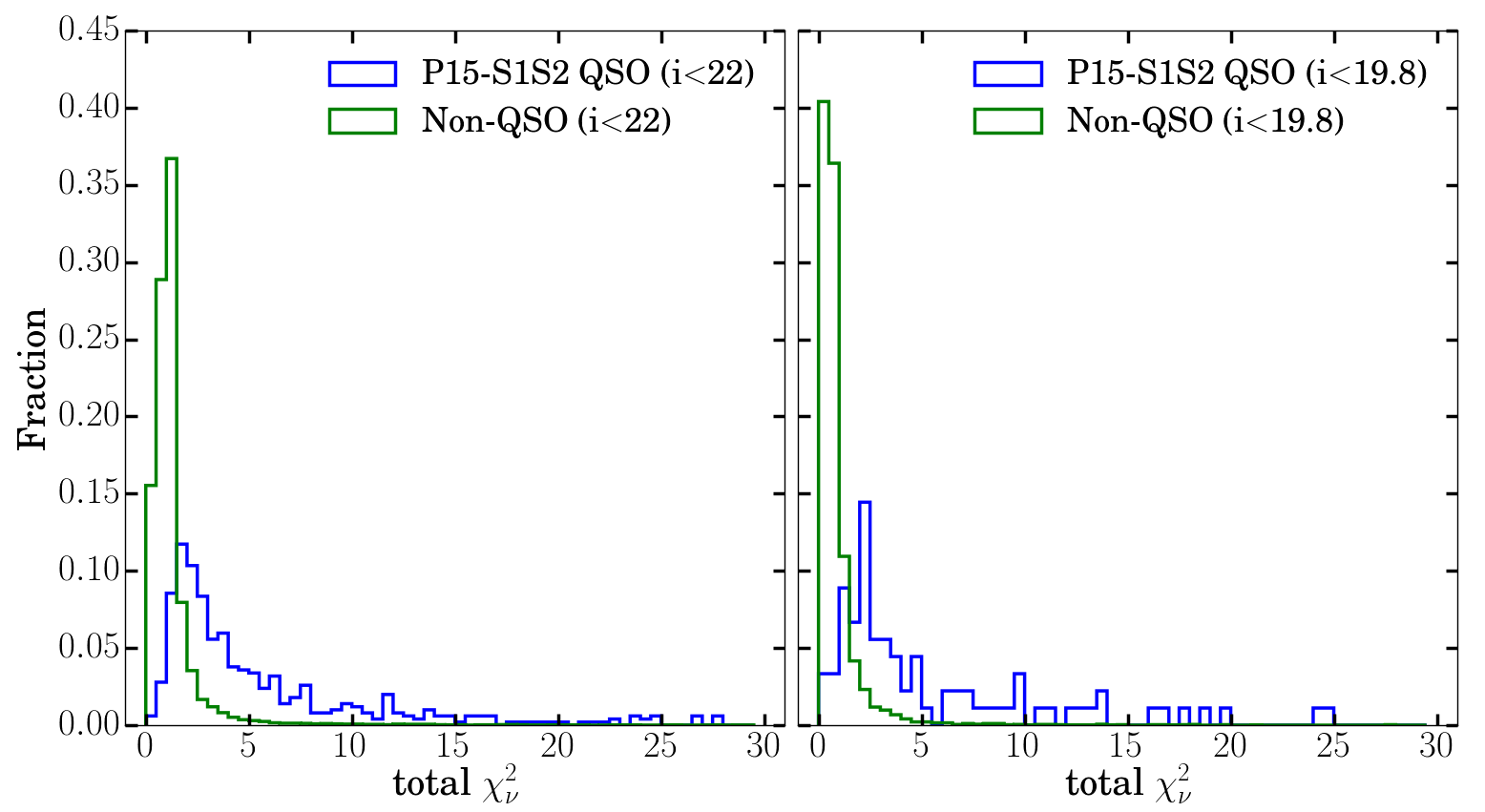}
    \label{fig:chi2}
    \caption{Reduced-$\chi^{2}$ distribution over DES bands with enough epochs for the P15-S1S2 quasars and point source non-quasars at the faint and bright magnitude limits. While the non-quasars are well-fit by a constant magnitude model, the quasars show larger deviations from such a model. }
\end{figure*}

Our variability criterion uses the chi-squared integrated probability $P(X_{\nu}^{2}\geq\chi^{2})$ to reject the null hypothesis, where $\nu = \sum N_{k}-\sum k$ is the number of degrees of freedom and $X_{\nu}^{2}$ is the chi-squared distribution with $\nu$ degrees of freedom. The sources are required to have $N_{k} \geq 3$ and $\sum N_{k} \geq 6$. In other words, a source needs to have at least six epochs of data if only one band is available or at least three epochs of data in at least two bands. After the application of the shape, flag, and photometric cuts, 503 (90) quasars and 17,539 (6,656) other point sources satisfied this condition at $i < 22$ mag ($i < 19.8$ mag). Figure \ref{fig:chi2} shows their reduced-$\chi^{2}$ distributions. 

\begin{figure}[tb]
	\centering
    \includegraphics[width=0.5\textwidth]{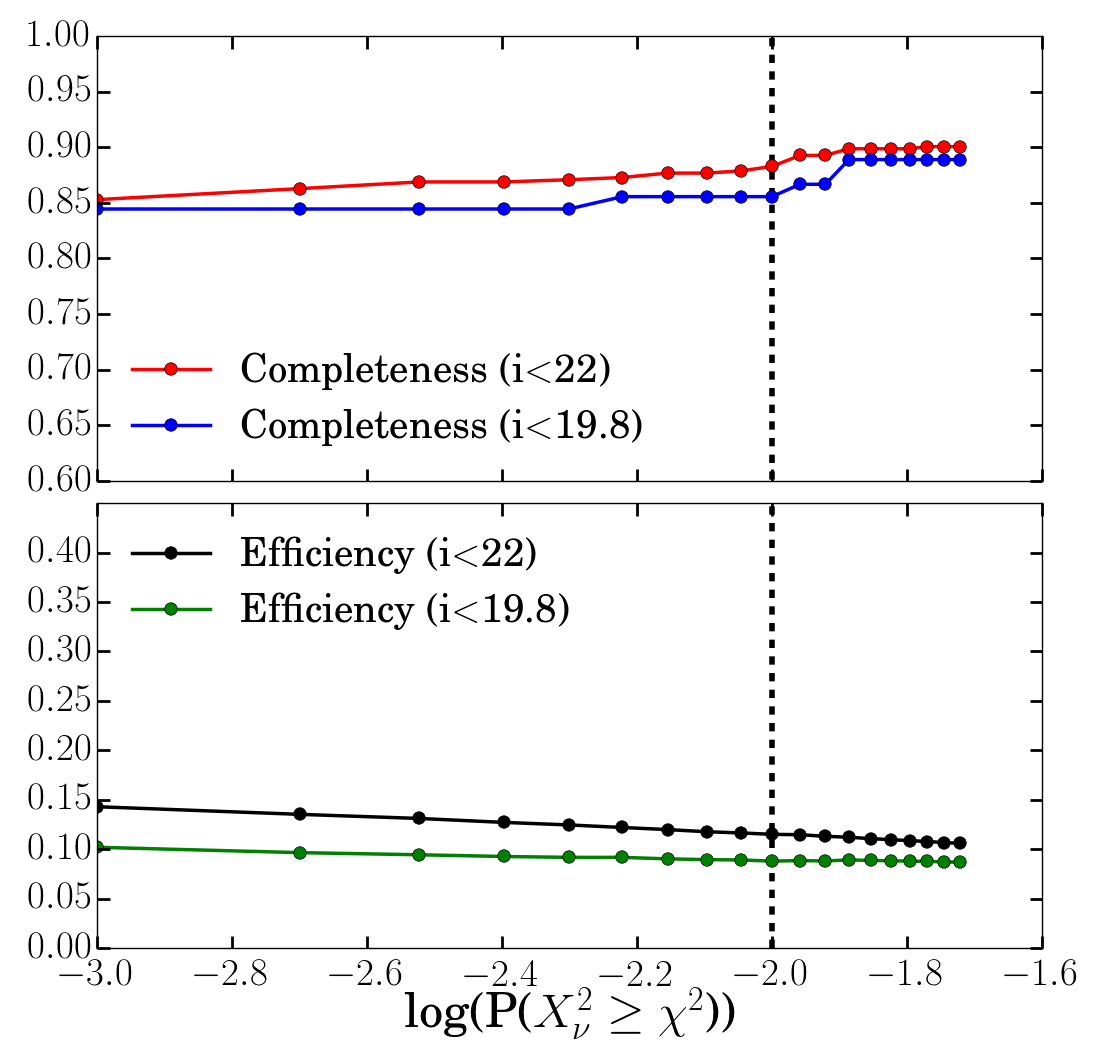}
    \label{fig:variability}
    \caption{Completeness and efficiency of variability selection $P(X_{\nu}^{2}\geq\chi^{2})$ $>$ 0.01 for the P15-S1S2 Quasar Catalog, where $P(X_{\nu}^{2}\geq\chi^{2})$ is the probability to reject the null hypothesis that sources are non-variable. The dotted line at log($P(X_{\nu}^{2}\geq\chi^{2})$)=$-$2.00 corresponds to rejecting the null hypothesis at 99\% confidence. }
\end{figure}

We chose a threshold $P(X_{\nu}^{2}\geq\chi^{2})$ value of 0.01 as our variability selection cut, which corresponds to rejection of the null hypothesis with 99\% significance. There are 444 (77) quasars and 3,418 (799) non-quasars that pass the variability selection criterion with $P(X_{\nu}^{2}\geq\chi^{2})$ $< $ 0.01, resulting in an efficiency of 11.4\% (8.8\%) and a completeness of 88\% (86\%). Figure \ref{fig:variability} shows the efficiency and completeness as a function of the \textit{P}-value for the P15-S1S2 quasars. The low efficiencies are likely due to the short baseline of our current light curves combined with contamination from stars which have a higher surface density and show low-level variability. Since our variability selection is simply based on the $\chi^{2}$ bound for fitting a constant magnitude, any sufficiently variable source will pass the cut. More sophisticated statistical models require longer observational baselines, but can differentiate between quasar and stellar variability. The average number of epochs in the present variability sample is $\sim$13 in \textit{gri} bands and $\sim$40 in \textit{z} band spread over only half a year. This is much smaller than the $\sim$60 epochs spanning over $\sim$6 years in the SDSS Stripe 82 region \citep{Sesar2007,Bramich2008}. With the full 5-year DES data, the number of epochs will increase by a factor of 5 and the probability a quasar has varied significantly will increase due to the longer time baseline. This will improve the performance of the variability selection, as well as enabling the use of more sophisticated quasar variability models.


We investigated the performance of combinations of $W1W2$ color, $giW1$ color selection, and \texttt{XDQSOz} methods with variability. The combinations $W1W2$+variability, $giW1$+variability, and \texttt{XDQSOz}+variability give completenesses of 88\% (81\%), 86\% (78\%), and 85\% (83\%) and efficiencies of 63\% (83\%), 60\% (80\%), and 68\% (83\%), respectively, for the total (bright) sample. Compared to color selection, \texttt{XDQSOz} selection or variability selection alone, these combinations produce higher selection efficiency due to fewer false positives, with a slight reduction in completeness.  


\section{The OzDES Quasar Catalog}
\label{sec:ozdes_qso_cat}
The Australian Dark Energy Survey (OzDES; \citealp{Yuan2015}) is a spectroscopic survey of the DES supernova fields using the AAOmega spectrograph \citep{smith2004} on the Anglo-Australian Telescope (AAT). The field of view of the AAT multi-object fibre-positioning system \citep{Lewis2002} is well-matched to DECam, making the AAT a well-suited spectroscopic follow-up instrument for DES. 

OzDES commenced observations at the same time as DES in 2013B and aims to measure redshifts for thousands of Type Ia supernova host galaxies and black hole masses for hundreds of active galactic nuclei (AGN) and quasars \citep{King2015} using reverberation mapping \citep{Blandford1982,Peterson1993}. Objects that range in brightness from $m_{r}$ $\sim$ 17 mag to $m_{r}$ $\sim$ 25 mag are selected for spectroscopy via a variety of criteria, such as supernova candidates, quasar candidates, galaxy cluster members, and photometric redshift calibration candidates (see \citealp{Yuan2015} for further details). Under most circumstances, an object is repeatedly observed until a redshift is obtained. The observing targets and results of OzDES observations are compiled into a spectroscopic catalog known as the Global Redshift Catalog (GRC), which also includes redshifts from other spectroscopic surveys such as 6dF \citep{Jones2004}, SDSS \citep{York2000}, and VVDS \citep{LF2005}. The GRC is updated after every OzDES observing season. In this work, we used the February 2016 version of the OzDES GRC. 

Here we present the OzDES Quasar Catalog of 1,263 OzDES sources with M$_{i}$ $< -22$ mag and $i$ $<$ 22 mag that are spectroscopically confirmed quasars. We constructed the catalog as follows. As the GRC does not provide photometric information, we cross-matched the GRC with the DES supernova field catalog \textit{Y1A1\_COADD\_OBJECTS\_DFULL} (see \S\ref{subsec:des}). Sources within 1.1$\arcdeg$ of the ten supernova field centers (see Table \ref{tab:des_sn}) and with \textit{i} $<$ 22 mag are matched using a matching radius of 0\farcs5. We used k-corrections from \cite{Richards2006A} and  the astropy cosmology package\footnote{\url{ http://www.astropy.org}} to calculate the absolute $i$-band magnitude M$_{i}$ of the OzDES sources for a flat $\Lambda$CDM cosmology with $H_{0}$ = 70 km s$^{-1}$Mpc$^{-1}$ and $\Omega_{0}$=0.3. Specifically, as the k-corrections in \cite{Richards2006A} are for z=2, we applied the offset $M_{i}(z=0)$ = $M_{i}(z=2)$ + 0.596 in accordance with Eq. (1) of \cite{Richards2006A}. We narrowed down the sample to sources with ``good'' OzDES redshifts (quality flags 3 or 4 in the GRC). Finally, we visually inspected the spectra of sources that have M$_{i}$ $< -22$ mag, which are roughly 2,200 total objects. We looked for the presence of broad emission lines, such as Ly${\alpha}$, C IV 1548, CIII 1909, and Mg II 2798, and ensured that the line identifications are consistent with the OzDES redshifts. For sources with no clear lines ($\sim$ 40\% of the visually inspected sources), we re-examined their spectra and redshifts using the OzDES redshifting software MARZ \citep{Hinton2016}. Only sources that passed our visual inspection are included in the catalog. 

\begin{figure*}[tb]
\centering
\includegraphics[width=0.9\textwidth]{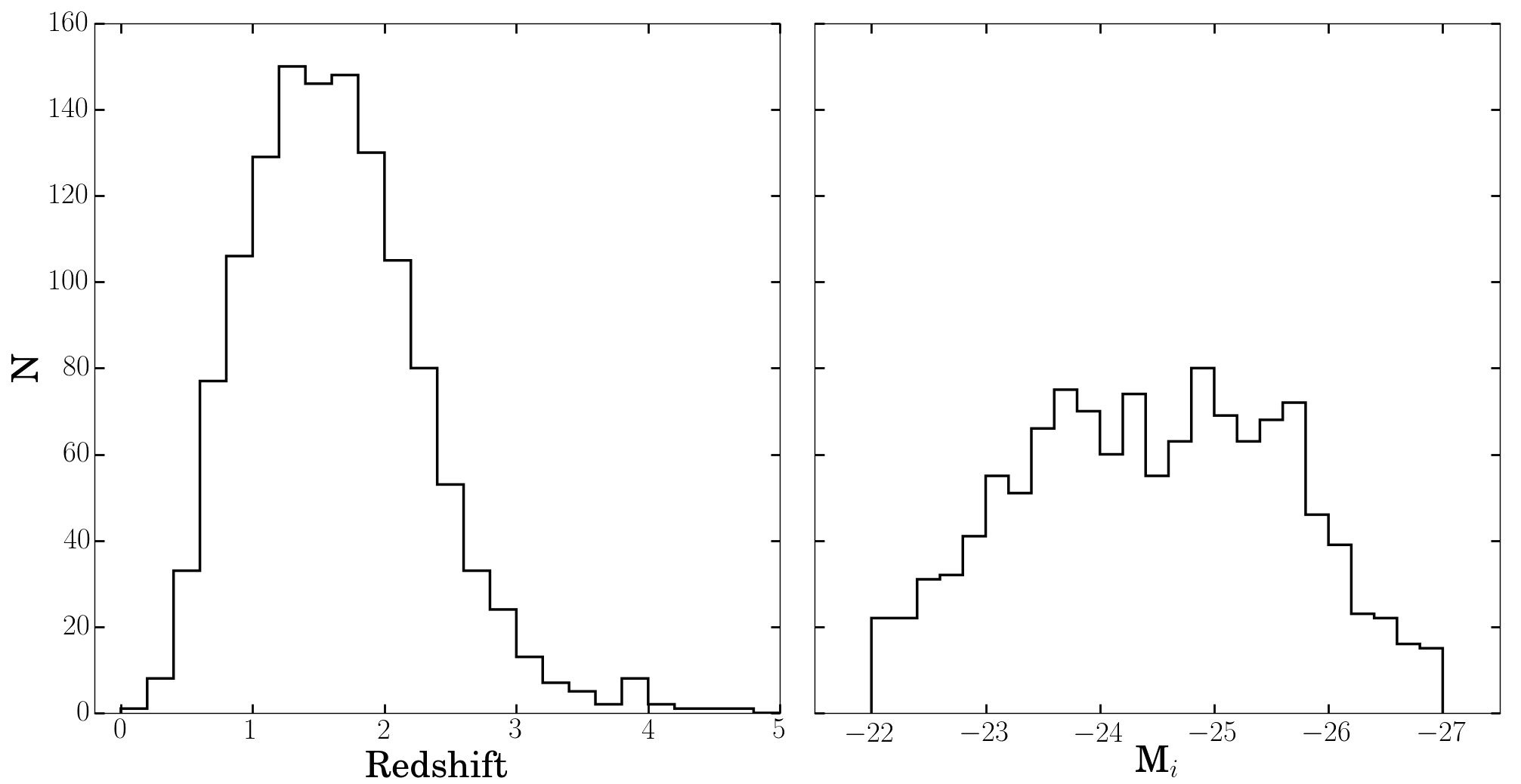}
\label{fig:ozdes_Miz}
\caption{Absolute $i$-band magnitude M$_{i}$ and redshift distributions of the OzDES Quasar Catalog. The quasars are selected to have M$_{i}$ $< -22$ mag and show clear broad/narrow emission/absorption lines in their OzDES spectra. }
\end{figure*}

The OzDES Quasar Catalog consists of 1,263 quasars brighter than \textit{i}=22 mag, corresponding to a quasar surface density of $\sim$ 42 deg$^{-2}$. The quasars discovered by OzDES include a recently discovered post-starburst broad absorption line (BAL) quasar at z=0.65 (see \citealp{Mudd2016}). The catalog includes multi-wavelength photometry from DES, VHS, and WISE is provided as part of the catalog. Table \ref{tab:ozdes_qso_catalog} describes the catalog and the full table is available in the electronic version of the Journal. Figure \ref{fig:ozdes_Miz} shows the absolute $i$-band magnitude and redshift distributions of the quasars. 

The OzDES Quasar Catalog is neither homogeneously selected nor complete. The OzDES selection codes in Table \ref{tab:ozdes_qso_catalog} note how the quasars were initially selected for OzDES observations. For instance, the selection codes 1, 2, and 16 refer to previously known quasars from DES Science Verification (SV) observations \citep{Banerji2015}, the VVDS survey, and NED\footnote{\url{https://ned.ipac.caltech.edu/}}, respectively, while quasars with selection code 0 are mostly identified as transients and DES photo-$z$ calibration targets. Selection code 4 refers to objects that were best matched to a quasar template with LePhare photo-$z$ estimation tool \citep{Arnouts1999}. Target selection with LePhare only used the stellar and quasar templates, as the input catalog is limited to point-like sources. Selection codes 8, 32, and 64 are color-selected quasars using the color cuts described in \S\ref{subsec:giw1_selection} and \S\ref{subsec:gik_selection} for sources that satisfied a shape cut. Finally, quasars with selection codes 128 and 256 are selected using SDSS color cuts from \cite{Ross2012}: $g-r$ vs. $r-i$ and $r-i$ vs. $i-z$ in the case of code 128 and $u-g$ vs. $g-r$ in the case of code 256 (a limited amount of $u$-band data was obtained during DES Science Verification in 2012B). We aim to improve the completeness of the OzDES Quasar Catalog as the DES and OzDES observations continue. 

\begin{table*}[t]
\small
\centering
\smallskip
\label{tab:ozdes_qso_catalog}
\caption{OzDES Quasar Catalog}
\begin{threeparttable}
  \begin{tabular}{l l}
	\toprule
	Column Name & Description\\[0.5ex]
    \midrule
ID &  OzDES Global Redshift Catalog (GRC) ID\\[0.5ex]
DES coadd\_objects\_id &  DES ID from the \textit{Y1A1\_COADD\_OBJECTS\_DFULL} catalog\\[0.5ex]
RA, DEC &  Right Ascension, Declination\\[0.5ex]
REDSHIFT &  spectrosopic redshift from OZDES\\[0.5ex]
ABS\_I\_MAG &  absolute $i$-band magnitude calculated using the OzDES redshift\\[0.5ex]
DES wavgcalib\_mag\_psf\_[g,r,i,z,Y] &  DES magnitudes (AB) (see \S\ref{subsec:des})\\[0.5ex]
DES wavg\_mag\_err\_[g,r,i,z,Y] & DES magnitude errors\\[0.5ex]
DES flags\_[g,r,i,z,Y] &  DES flags from SExtractor\\[0.5ex]
DES wavg\_spread\_model\_[g,r,i,z,Y] & DES spread\_model from SExtractor (see \S\ref{sec:shape_selection})\\[0.5ex]
DES wavg\_spreaderr\_model\_[g,r,i,z,Y] & DES spread\_model errors\\[0.5ex]
WISE w1mpro, w2mpro &  WISE magnitudes (Vega) from the \textit{ALLWISE} catalog (see \S\ref{subsec:wise})\\[0.5ex]
WISE w1sigmpro, w2sigmpro &  WISE magnitude errors\\[0.5ex]
VHS [y,j,h,ks]AperMag3 &  VHS magnitudes (Vega) from the DR3 \textit{vhsSource} catalog (see \S\ref{subsec:vhs})\\[0.5ex]
VHS [y,j,h,ks]AperMag3Err &  VHS magnitude errors\\[0.5ex]
OzDES Selection Code & 0: Not targeted as a quasar candidate\\[0.5ex]
& 1: DES SV targets\\[0.5ex]
& 2: VVDS quasars\\[0.5ex]
& 4: QSO selection using the LePhare photo-$z$ estimation tool\\[0.5ex]
& 8: $i<$21.5 + $giK$ selection (see \S\ref{subsec:gik_selection}) + $K_{\rm Vega}$ $>$ 14 + $W1-W2$ $>$ 0.7\\[0.5ex]
& 16: NED quasars\\[0.5ex]
& 32: $giW1$ selection (see \S\ref{subsec:giw1_selection})\\[0.5ex]
& 64: Similar selection to code 8 but with a relaxed blue cut in $giK$\\[0.5ex]
& 128: High redshift selection using \textit{gri} and \textit{riz} colors\\[0.5ex]
& 256: \textit{u}-band selection\\[0.5ex]
	 \bottomrule
	\end{tabular}
 	\begin{tablenotes}
      \small
      \item Quasars with non-detections in DES bands, no WISE matches, or no VHS matches have entries of 99 for their magnitudes and errors. The naming schemes for the photometry follow the catalogs from which they are derived; see the relevant sections as indicated for more information. There are 44 columns in the catalog. 
    \end{tablenotes}
    \end{threeparttable}
\end{table*}

\section{Discussion and Summary}
\label{sec:discussion}
Large samples of quasars are needed to measure quasar luminosity functions and their evolution, particularly at lower luminosities, to estimate quasar host halo masses with weak lensing, and to study baryonic acoustic oscillations (BAO) with the Ly-$\alpha$ forest, most notably with the Dark Energy Spectroscopic Instrument (DESI)\footnote{http://desi.lbl.gov/tdr/}. We used the \cite{Peters2015} quasar catalog for the SDSS Stripe 82 region that overlaps the DES supernova fields, which we refer to as the P15-S1S2 Quasar Catalog, to analyze the completeness and efficiency of quasar color selection, probabilistic selection, variability selection, and combinations of colors and probabilistic selection methods with variability at the bright ($i$ $<$ 19.8 mag) and faint ($i$ $<$ 22 mag) magnitude limits.  We used only point sources that satisfy our data quality and photometric error cuts. The sources used to investigate each method also need to meet the distinct detection thresholds for each selection method because of their dependence on the external WISE and VHS catalogs. 

\begin{figure*}[tb]
	\centering
    \includegraphics[width=0.9\textwidth]{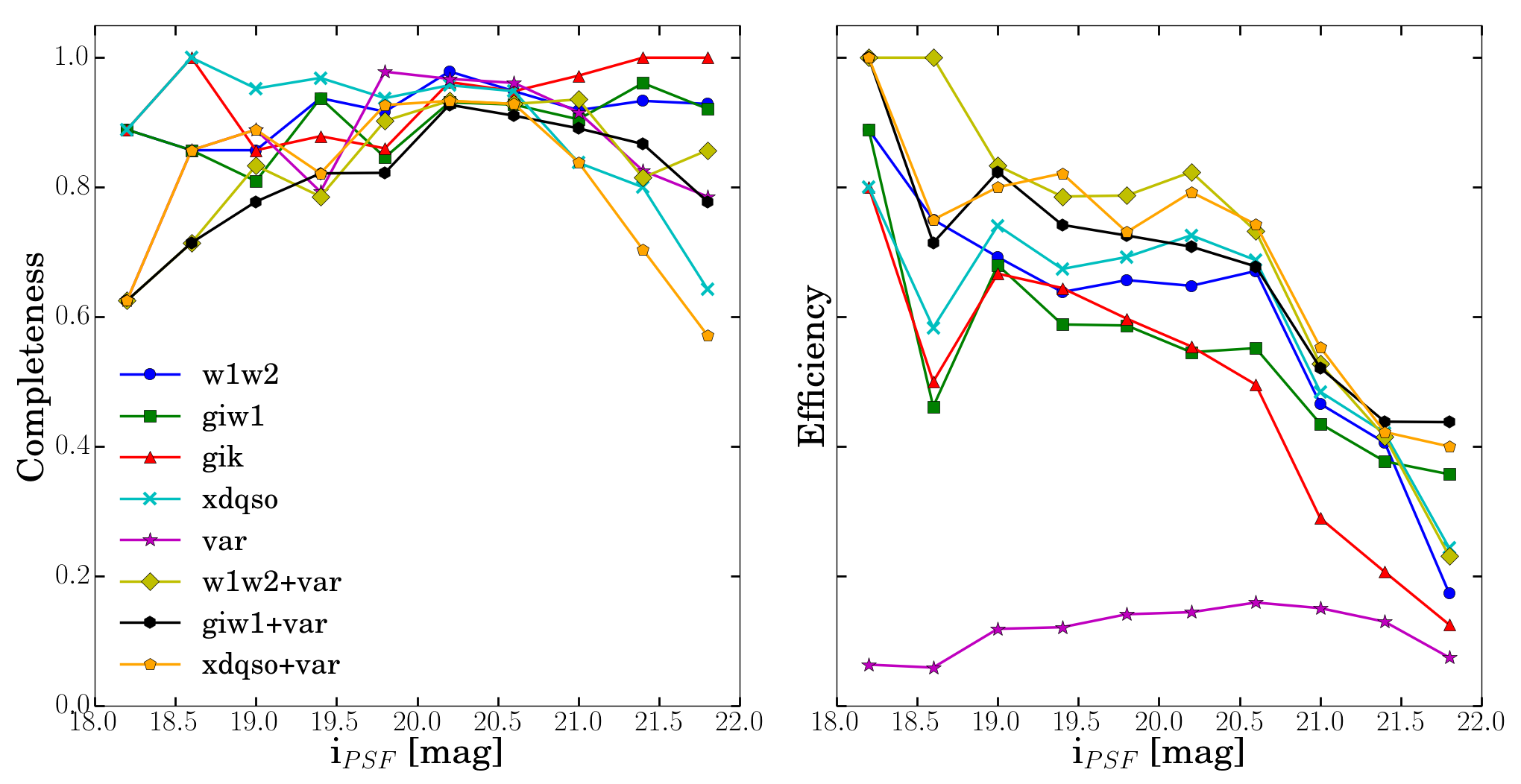}
    \label{fig:summary_peters}
    \caption{Completeness and efficiency as a function of $i$-band magnitude for the quasar selection methods investigated in this work with the P15-S1S2 Quasar Catalog. }
\end{figure*}

The results of our analyses are summarized in Table \ref{tab:summary} and Figure \ref{fig:summary_peters}. 
Figure \ref{fig:normalization} shows the fraction of sources that meet the selection-dependent thresholds (i.e $f_{det}$ from Table \ref{tab:summary}) for each selection method as a function of the DES $i$-band magnitude. The completeness and efficiency values both have statistical Poissonian uncertainties due to the actual number of quasars in this study, and systematic uncertainties due to cosmic variance in the quasar population and the characteristics of the DES data for the S1 and S2 fields. For a typical sample size of 300 (50) quasars in the faint (bright) samples used in this study, the statistical uncertainties are on the order of 8\% (18\% $-$ 22\%) and these serve as guidelines to compare the relative performance of the selection methods. These percentages correspond to one-sided, 90\% confidence limits. 

We calculated the surface densities of quasars and all candidates (quasars and non-quasars) selected by each selection method, also shown in Table \ref{tab:summary}. As the area of Stripe 82 is more precisely known than the overlap between DES S1 and S2 fields with Stripe 82, and partly because cosmic variance is smaller over a larger region, the surface densities are calculated using the overall P15 quasar surface densities, rather than using the sources in the DES S1 and S2 fields alone. For their ``good'' quasar candidates (\S\ref{sec:peters_cat}), the P15 catalog has a surface density $\Sigma_{\rm 22}$ = 132.79 deg$^{-2}$ at $i_{\rm SDSS}$ $<$ 22 mag and $\Sigma_{\rm 19.9}$ = 24.36 deg$^{-2}$ at $i_{\rm SDSS}$ $<$ 19.9 mag. The surface density of quasars $\Sigma_{\rm QSO}$ and all candidates $\Sigma_{\rm All}$ (quasars and non-quasars) selected by a selection method is then $\Sigma_{\rm QSO}$ = $\Sigma_{mag}$ $f_{det}$ $C$ and $\Sigma_{\rm All}$ = $\Sigma_{mag}$ $f_{det}$ $C$/$E$ = $\Sigma_{\rm QSO}$/$E$, where $\Sigma_{mag}$ refers to either $\Sigma_{22}$ or $\Sigma_{19.9}$, $f_{det}$ is from Table \ref{tab:summary}, $C$ is the completeness, and $E$ is the efficiency. The surface density of selected non-quasars is simply $\Sigma_{\rm All}$ $-$ $\Sigma_{\rm QSO}$.

While variability selection gives the highest surface density, the selection efficiency is one of the lowest and this results in a significant fraction of contaminants. Among individual selection methods, $giW1$ returns the highest quasar surface density, although it has a lower efficiency than $W1W2$ and \texttt{XDQSOz}. For combined selection methods, \texttt{XDQSOz}+variability is more efficient than $W1W2$+variability and $giW1$+variability but also returns lower quasar surface density. The $giW1$+variability selection gives the highest quasar surface density among the three combined selection methods. Taking both the surface densities of selected quasars and selection efficiencies into account, \texttt{XDQSOz} selection alone and a combination of color or \texttt{XDQSOz} with variability result in relatively high surface densities of quasars and a modest amount of contamination. The most significant drawback for both the $W1W2$ and \texttt{XDQSOz} methods is that the WISE data are only available for 55\% of the quasars with $i$ $<$ 22 mag, as both $W1$ and $W2$ detections are required. While the $giW1$ selection method has lower efficiency, it will produce a higher surface density of quasars because a $W2$ detection is not required. 



Since we used only photometry from the first year of DES operations, we employed a simple multi-band $\chi^{2}$ to detect variability. Variability selection based on more sophisticated quasar variability models, such as the damped random walk model, is deferred to the future when substantially more epochs will be available. The combination of the $W1W2$, $giW1$, or \texttt{XDQSOz} methods with variability improves the efficiency of quasar selection (fewer false positives) at slightly lower completeness. At the bright end, $W1W2$+variability and \texttt{XDQSOz}+variability can be applied in the DES supernova fields, but the fraction of sources with WISE detections will diminish for fainter sources. The $giW1$ or $giW1$+variability selection method can be used instead, particularly because they both return high quasar surface densities, and in the case of $giW1$+variability, with good selection efficiency. The depth of WISE data is less of an issue for the shallower fields. Although variability selection will not be as useful in these fields because of the small number of epochs, $W1W2$, $giW1$ or \texttt{XDQSOz} selection would be a good alternative. 

We also presented the OzDES Quasar Catalog of 1,263 spectroscopically-confirmed quasars in the 30 deg$^{2}$ DES Supernova fields brighter than \textit{i}=22 mag. The catalog includes all the quasars selected from the DES/OzDES reverberation mapping project with good quality OzDES redshifts, M$_{i}$ $<$ $-$22 mag, and visually confirmed emission and absorption lines. The OzDES Quasar Catalog is not homogeneous or complete, although its completeness will improve as the OzDES observations continue. 

\begin{table*}[tb]
  \centering
   \caption{Summary of quasar selection methods using the P15-S1S2 quasars.}
\label{tab:summary}
  \begin{threeparttable}
  \begin{tabular}{*{13}{l}}
    \toprule
    & \multicolumn{2}{c}{Completeness} & \multicolumn{2}{c}{Efficiency} 
      & \multicolumn{2}{c}{$C\times E$} & \multicolumn{2}{c}{$f_{det}$} 
      & \multicolumn{2}{c}{$\Sigma_{\rm QSO}$}(deg$^{-2}$) & \multicolumn{2}{c}{$\Sigma_{\rm All}$} (deg$^{-2}$)\\
    \cmidrule(lr){2-3}
    \cmidrule(lr){4-5}
    \cmidrule(lr){6-7}
    \cmidrule(lr){8-9}
    \cmidrule(lr){10-11}
    \cmidrule(lr){12-13}
    Selection & Bright & Total & Bright & Total & Bright & Total & Bright & Total & Bright & Total
    & Bright & Total\\
    &  &  &  &  &  &  &  &  & ($\Sigma_{\rm P15}$=24.4) & ($\Sigma_{\rm P15}$=132.8) &  & \\
    \midrule
    $W1W2$ & 0.91 & 0.93 & 0.71 & 0.54 & 0.65 & 0.50 & 0.96 & 0.55 & 21.3 & 67.9 & 30.0 & 125.8\\
    $giW1$ & 0.87 & 0.91 & 0.61 & 0.49 & 0.53 & 0.44 & 0.98 & 0.72 & 20.8 & 87.0 & 34.1 & 177.6\\
    $giK$  & 0.89 & 0.92 & 0.63 & 0.44 & 0.56 & 0.41 & 0.99 & 0.52 & 21.5 & 63.5 & 34.1 & 144.4\\
    \texttt{XDQSOz} & 0.92 & 0.90 & 0.73 & 0.60 & 0.67 & 0.54 & 0.96 & 0.55 & 21.5 & 65.7 & 29.5 & 109.6\\
    Variability & 0.86 & 0.88 & 0.088 & 0.11 & 0.075 & 0.10 & 0.86 & 0.90 & 18.0 & 105.2 & 204.7 & 956.1\\
    $W1W2$$+$Var & 0.81 & 0.88 & 0.83 & 0.63 & 0.68 & 0.55 & 0.82 & 0.49 & 16.2 & 57.3 & 19.5 & 90.9 \\
    $giW1$$+$Var & 0.78 & 0.86 & 0.80 & 0.60 & 0.63 & 0.51 & 0.84 & 0.65 & 16.0 & 74.2 & 20.0 & 123.7 \\    
    \texttt{XDQSOz}$+$Var & 0.83 & 0.85 & 0.83 & 0.68 & 0.68 & 0.57 &0.82& 0.49 & 16.6 & 55.3 & 20.0 & 81.3\\
    \bottomrule
  \end{tabular}
  \begin{tablenotes}
      \small
      \item The bright sample refers to sources with $i$ $<$ 19.8 mag and the total sample refers to sources with $i$ $<$ 22 mag. Completeness and efficiency are defined at the end of \S\ref{sec:peters_cat}. ``$C\times E$'' is the product of completeness and efficiency. ``$f_{det}$'' is the fraction of quasars that meet the selection-dependent detection threshold(s). This fraction is relative to all quasars that satisfy our point source, flags, and photometric errors cuts. Since both the $W1W2$ and \texttt{XDQSOz} methods have the same DES and WISE detection requirements, their $f_{det}$ values are the same. The statistical uncertainties in the number of quasars for a typical sample size of 300 (50) in the faint (bright) sample used in this study are on the order of 8\% (18\% $-$22\%). For more discussion, see \S\ref{sec:discussion}. The last four columns show the surface densities of selected quasars ($\Sigma_{\rm QSO}$) and all candidates ($\Sigma_{\rm All}$) for each method. The surface densities are calculated based on the overall P15 quasar surface density ($\Sigma_{\rm P15}$), $f_{det}$, and completeness/efficiency as detailed in \S\ref{sec:discussion}. The surface density of the selected non-quasars is simply $\Sigma_{\rm All}$ $-$ $\Sigma_{\rm QSO}$. 
    \end{tablenotes}
  \end{threeparttable}
\end{table*}

\section*{Acknowledgment}
Funding for the DES Projects has been provided by the U.S. Department of Energy, the U.S. National Science Foundation, the Ministry of Science and Education of Spain, 
the Science and Technology Facilities Council of the United Kingdom, the Higher Education Funding Council for England, the National Center for Supercomputing Applications at the University of Illinois at Urbana-Champaign, the Kavli Institute of Cosmological Physics at the University of Chicago, the Center for Cosmology and Astro-Particle Physics at the Ohio State University, the Mitchell Institute for Fundamental Physics and Astronomy at Texas A\&M University, Financiadora de Estudos e Projetos, Funda{\c c}{\~a}o Carlos Chagas Filho de Amparo {\`a} Pesquisa do Estado do Rio de Janeiro, Conselho Nacional de Desenvolvimento Cient{\'i}fico e Tecnol{\'o}gico and the Minist{\'e}rio da Ci{\^e}ncia, Tecnologia e Inova{\c c}{\~a}o, the Deutsche Forschungsgemeinschaft and the Collaborating Institutions in the Dark Energy Survey. 

The Collaborating Institutions are Argonne National Laboratory, the University of California at Santa Cruz, the University of Cambridge, Centro de Investigaciones Energ{\'e}ticas, Medioambientales y Tecnol{\'o}gicas-Madrid, the University of Chicago, University College London, the DES-Brazil Consortium, the University of Edinburgh, 
the Eidgen{\"o}ssische Technische Hochschule (ETH) Z{\"u}rich, 
Fermi National Accelerator Laboratory, the University of Illinois at Urbana-Champaign, the Institut de Ci{\`e}ncies de l'Espai (IEEC/CSIC), 
the Institut de F{\'i}sica d'Altes Energies, Lawrence Berkeley National Laboratory, the Ludwig-Maximilians Universit{\"a}t M{\"u}nchen and the associated Excellence Cluster Universe, the University of Michigan, the National Optical Astronomy Observatory, the University of Nottingham, The Ohio State University, the University of Pennsylvania, the University of Portsmouth, SLAC National Accelerator Laboratory, Stanford University, the University of Sussex, Texas A\&M University, and the OzDES Membership Consortium.

The DES data management system is supported by the National Science Foundation under Grant Number AST-1138766. The DES participants from Spanish institutions are partially supported by MINECO under grants AYA2012-39559, ESP2013-48274, FPA2013-47986, and Centro de Excelencia Severo Ochoa SEV-2012-0234. Research leading to these results has received funding from the European Research Council under the European Union’s Seventh Framework Programme (FP7/2007-2013) including ERC grant agreements 240672, 291329, and 306478.

This work is based in part on data obtained at the Australian Astronomical Observatory through program A/2013B/012. This research made use of Astropy, a community-developed core Python package for Astronomy (Astropy Collaboration, 2013).

\section*{Affiliations}
\noindent
$^{1}$ Department of Astronomy, The Ohio State University, Columbus, OH 43210, USA \\*
$^{2}$ Center for Cosmology and Astro-Particle Physics, The Ohio State University, Columbus, OH 43210, USA \\*
$^{3}$ Institute of Astronomy, University of Cambridge, Madingley Road, Cambridge CB3 0HA, UK \\*
$^{4}$ Kavli Institute for Cosmology, University of Cambridge, Madingley Road, Cambridge CB3 0HA, UK \\*
$^{5}$ Australian Astronomical Observatory, North Ryde, NSW 2113, Australia \\*
$^{6}$ School of Mathematics and Physics, University of Queensland,  Brisbane, QLD 4072, Australia \\*
$^{7}$ The Research School of Astronomy and Astrophysics, Australian National University, ACT 2601, Australia \\*
$^{8}$ Centre for Astrophysics \& Supercomputing, Swinburne University of Technology, Victoria 3122, Australia \\*
$^{9}$ Fermi National Accelerator Laboratory, P. O. Box 500, Batavia, IL 60510, USA \\*
$^{10}$ Research School of Astronomy and Astrophysics, The Australian National University, Canberra, ACT 2611, Australia \\*
$^{11}$ INAF- Osservatorio Astronomico di Torino - Strada Osservatorio 20, Pino Torinese, 10020, Italy \\*
$^{12}$ School of Physics and Astronomy, University of Southampton, Southampton, SO17 1BJ, UK \\*
$^{13}$ Sydney Institute for Astronomy, School of Physics, A28, The University of Sydney, NSW 2006, Australia \\*
$^{14}$ Cerro Tololo Inter-American Observatory, National Optical Astronomy Observatory, Casilla 603, La Serena, Chile \\*
$^{15}$ Department of Physics \& Astronomy, University College London, Gower Street, London, WC1E 6BT, UK \\*
$^{16}$ Department of Physics and Electronics, Rhodes University, PO Box 94, Grahamstown, 6140, South Africa \\*
$^{17}$ CNRS, UMR 7095, Institut d'Astrophysique de Paris, F-75014, Paris, France \\*
$^{18}$ Sorbonne Universit\'es, UPMC Univ Paris 06, UMR 7095, Institut d'Astrophysique de Paris, F-75014, Paris, France \\*
$^{19}$ Laborat\'orio Interinstitucional de e-Astronomia - LIneA, Rua Gal. Jos\'e Cristino 77, Rio de Janeiro, RJ - 20921-400, Brazil \\*
$^{20}$ Observat\'orio Nacional, Rua Gal. Jos\'e Cristino 77, Rio de Janeiro, RJ - 20921-400, Brazil \\*
$^{21}$ Department of Astronomy, University of Illinois, 1002 W. Green Street, Urbana, IL 61801, USA \\*
$^{22}$ National Center for Supercomputing Applications, 1205 West Clark St., Urbana, IL 61801, USA \\*
$^{23}$ Institut de Ci\`encies de l'Espai, IEEC-CSIC, Campus UAB, Carrer de Can Magrans, s/n,  08193 Bellaterra, Barcelona, Spain \\*
$^{24}$ Institut de F\'{\i}sica d'Altes Energies (IFAE), The Barcelona Institute of Science and Technology, Campus UAB, 08193 Bellaterra (Barcelona) Spain \\*
$^{25}$ Kavli Institute for Particle Astrophysics \& Cosmology, P. O. Box 2450, Stanford University, Stanford, CA 94305, USA \\*
$^{26}$ George P. and Cynthia Woods Mitchell Institute for Fundamental Physics and Astronomy, and Department of Physics and Astronomy, Texas A\&M University, College Station, TX 77843,  USA \\*
$^{27}$ Department of Physics, IIT Hyderabad, Kandi, Telangana 502285, India \\*
$^{28}$ Jet Propulsion Laboratory, California Institute of Technology, 4800 Oak Grove Dr., Pasadena, CA 91109, USA \\*
$^{29}$ Department of Astronomy, University of Michigan, Ann Arbor, MI 48109, USA \\*
$^{30}$ Department of Physics, University of Michigan, Ann Arbor, MI 48109, USA \\*
$^{31}$ Kavli Institute for Cosmological Physics, University of Chicago, Chicago, IL 60637, USA \\*
$^{32}$ Instituto de F\'sica Te\'rica IFT-UAM/CSIC, Universidad Aut\'noma de Madrid, Cantoblanco 28049, Madrid, Spain \\*
$^{33}$ Department of Astronomy, University of California, Berkeley,  501 Campbell Hall, Berkeley, CA 94720, USA \\*
$^{34}$ Lawrence Berkeley National Laboratory, 1 Cyclotron Road, Berkeley, CA 94720, USA \\*
$^{35}$ SLAC National Accelerator Laboratory, Menlo Park, CA 94025, USA \\*
$^{36}$ Department of Physics, The Ohio State University, Columbus, OH 43210, USA \\*
$^{37}$ Astronomy Department, University of Washington, Box 351580, Seattle, WA 98195, USA \\*
$^{38}$ Departamento de F\'{\i}sica Matem\'atica,  Instituto de F\'{\i}sica, Universidade de S\~ao Paulo,  CP 66318, CEP 05314-970, S\~ao Paulo, SP,  Brazil \\*
$^{39}$ Instituci\'o Catalana de Recerca i Estudis Avan\c{c}ats, E-08010 Barcelona, Spain \\*
$^{40}$ Institute of Cosmology \& Gravitation, University of Portsmouth, Portsmouth, PO1 3FX, UK \\*
$^{41}$ Department of Physics and Astronomy, Pevensey Building, University of Sussex, Brighton, BN1 9QH, UK \\*
$^{42}$ Centro de Investigaciones Energ\'eticas, Medioambientales y Tecnol\'ogicas (CIEMAT), Madrid, Spain \\*
$^{43}$ Instituto de F\'\i sica, UFRGS, Caixa Postal 15051, Porto Alegre, RS - 91501-970, Brazil \\*
$^{44}$ Universidade Federal do ABC, Centro de Ci\^encias Naturais e Humanas, Av. dos Estados, 5001, Santo Andr\'e, SP, Brazil, 09210-580 \\*
$^{45}$ Universidade Federal do ABC, Centro de Ci\^encias Naturais e Humanas, Av. dos Estados, 5001, Santo Andr\'e, SP, Brazil, 09210-580 \\*
$^{46}$ Computer Science and Mathematics Division, Oak Ridge National Laboratory, Oak Ridge, TN 37831


\end{document}